\documentclass[]{article}
\usepackage{multicol}
\usepackage{apj}
\newcommand {\h} {$h^{-1} \, Mpc \,$}
\newcommand {\ks} {$km~s^{-1} \;$}

\begin{document}

\vspace{15mm}
\begin{center}
\uppercase{Optical substructures in 48 Galaxy Clusters:\\
 New  Insights from a Multi-Scale Analysis}\\
\vspace*{1.5ex}
{\sc M. Girardi$^{1,2}$, E. Escalera$^{1,2,3}$, D. Fadda$^{1,2}$, G.
 Giuricin$^{1,2}$, F. Mardirossian$^{1,2,4}$ and M. Mezzetti$^{1,2}$}\\
\vspace*{1.ex}
{\small 
$^1$Dipartimento di Astronomia, Universit\`{a} degli Studi di Trieste, \\
$^2$SISSA, via Beirut 4, 34013 - Trieste, Italy\\ 
$^3$Observatoire de Marseille, Place Le Verrier, F-13248, Marseille, C\'edex 4, France \\
$^4$Osservatorio Astronomico di Trieste, Italy\\
E-mail:
girardi@sissa.it; escalera@obmara.cnrs-mrs.fr;
giuricin@sissa.it; fadda@sissa.it; mezzetti@sissa.it}
\end{center}
\vspace*{-6pt}

\begin{abstract}

We analyze the presence of substructures in a set of 48
galaxy clusters, by using galaxy positions and redshifts.
The data are taken from literature sources,
with the addition of some new data provided by recent observations of
galaxy clusters.

We use a multi-scale analysis which couples kinematical estimators
with the wavelet transform.  With respect to previous works, we
introduce three new kinematical estimators.  These estimators
parameterize the departures of the local means and/or local
dispersions of the measured radial velocities with respect to their
global values for the environment.

We classify the analyzed clusters as unimodal, bimodal and complex
systems. 
We find that $\sim 14\%$ of our clusters are strongly
substructured (i.e. they are bimodal or complex) and that 
$\sim 24\%$ of the remaining unimodal clusters
contain substructures at small scales. 
Thus, in substantial agreement
with previous studies, about one third of clusters show
substructures.

We find that the presence of substructures in unimodal clusters does
not affect the estimates of both velocity dispersions and virial
masses. Moreover, the galaxy velocity dispersion is generally in good
agreement with the X-ray temperature, according to the expectations of
the standard isothermal model for galaxy clusters. These facts suggest
that unimodal clusters, which are the most frequent cases in the
nearby Universe, are not too far from a status of dynamical
equilibrium.

On the contrary, the estimates of velocity dispersions and masses for
some bimodal or complex clusters strongly depend on whether they are
treated as single systems or as sums of different clumps.  In these
cases the X-ray temperature and the velocity dispersion may be
very different.\\

\vspace*{-6pt}
\noindent
{\em Subject headings: } galaxies: clusters: general -- galaxies: clustering --
 cosmology: observations -- methods: data analysis

\end{abstract}

\begin{multicols}{2}
\section{Introduction}

The recent literature has provided firm evidence of the presence of
substructures in galaxy clusters (see West 1994, and references
therein).  Indeed, when the amount of data grows and the techniques of
analysis are improved, the clusters show more and more substructures
(see, e.g., recent results on the Coma cluster by Biviano et al. 1996).

The effect of substructures on cluster kinematics and dynamics is
widely studied in the literature.
The presence of substructures could make the galaxy velocity
distributions deviate from Gaussian ones (Bird \& Beers 1993;
Zabludoff, Franx, \& Geller 1993).  Beers \& Tonry (1986) suggested
that the constant density cores of clusters are actually due to the
presence of central substructures.  Substructures could also cause the
observed significant velocity offsets of cD galaxies with respect to
other cluster members (Sharples, Ellis, \& Gray 1988;
Hill et al. 1988).  

The presence of substructures could lead to over- or underestimates of
the galaxy velocity dispersions (e.g. Fitchett 1988), to overestimates
of the cluster mass (Pinkney et al. 1996), and could also modify the
velocity dispersion profile in the central cluster region (Fitchett \&
Webster 1987).  In particular, these effects are suggested as causing
the disagreement between the observed velocity dispersion of galaxies
and X-ray temperature of hot gas (Edge \& Stewart 1991b).  On the
other hand, collisions of subclusters can enhance the X-ray
temperature (e.g. Briel \& Henry 1994; Zabludoff \& Zaritsky 1995).

Also numerical simulations show that both galaxy velocity dispersion
and gas temperature increase during a phase of cluster merging (see
e.g.  Evrard 1990; Schindler \& Boehringer 1993; Schindler \& Mueller
1993; Roettiger, Burns, \& Loken 1993; Burns et al. 1995).  Therefore,
in cases of strong substructures, e.g. close bimodal clusters, both
galaxy velocity dispersion and gas X-ray temperature may be bad
measures of the cluster potential since the cluster may be very far
from a status of dynamical equilibrium.

The situation is less clear for clusters with small substructures,
which are the most frequent cases and thus the most important ones in
statistical analyses, e.g. in the observational distribution function
of cluster
masses, since, to evaluate cluster masses, a status of dynamical
equilibrium is generally assumed (e.g.  Bahcall \& Cen 1993; Biviano
et al. 1993). Recent results from numerical simulations suggest that,
on average, clusters could be approximately in dynamical equilibrium
within a central region (e.g. Tormen, Bouchet, \& White 1996).
From the observational point of view, partially contradictory
conclusions are reached by two recent works - based on large cluster
samples - which look for substructures by using galaxy positions and
redshifts (Escalera et al.  1994, hereafter E94; Bird 1995, hereafter
B95).

  By examining 16 clusters, E94 found that the sum of virial masses of
gravitationally bounded internal structures is generally close to the
total virial mass of the main cluster.  On the contrary, B95 found
that the correction for the presence of substructures appreciably
affects the masses of 25 clusters with a dominant galaxy, the effect
being mainly due to a reduction of the mean galaxy separation.
However, B95 found that the correction for substructures is not
significant if the cluster mass is computed within the virialization
radius (see B95) rather than within an Abell radius.  Both works agree
in claiming that the velocity dispersion is not strongly biased by the
presence of substructures.

Another critical question concerns the survival time of a substructure
within the cluster (see e.g.  Gonzales-Casado, Mamon, \&
Salvador-Sol\'e 1994), which is essential for constraining the critical
density of the Universe by using the frequency of substructures (Lacey
\& Cole 1993; Ueda et al.  1995). The poor knowledge of the frequency,
nature and origin of substructures makes the problem more
difficult.

The availability  of a large amount of new redshifts for cluster galaxies
(e.g.  Katgert et al.  1996) allows us to better investigate 
cluster structures.  The redshift information greatly alleviates the
problems induced by the presence of cluster interlopers and/or cluster
overlapping, problems which are always present in two-dimensional
analyses. With respect to studies based on X-ray data, an
optical analysis may have the advantages of allowing a
three-dimensional analysis, of identifying the galaxies belonging
to different subclumps, and of investigating the outer cluster regions
of low X-ray surface brightness.  On the other hand, because of the
still small number of measured galaxy redshifts we need very refined
techniques for substructure analyses.

For instance, a suitable technique is wavelet analysis, which can
be performed on optical, X-ray (Slezak et al. 1994), and 
radio data (Grebenev et al.  1995).  The wavelet analysis was first
applied to astrophysics as a two-dimensional technique by Slezak et
al. (1990).  Subsequently, Escalera \& Mazure (1992) improved the
technique by coupling it to redshift information. In this paper we
describe a further improvement in order to better detect substructures
in galaxy clusters.

The identification of galaxies involved in structures allows us both
to collect information concerning the substructures themselves and to
analyze their kinematical and dynamical effect on clusters.  The
difficulty in these analyses arises from the possible presence, also
in virialized systems, of velocity anisotropies in galaxy orbits, which
makes it difficult to deproject l.o.s. (i.e. line of sight) 
galaxy velocities.  These
problems are taken into account in our analyses, as well as in our
substructure detection.

In \S~2 we describe the main dataset used in this work.  \S~3 is a
description of our method.  In \S~4 we display the main results of the
structure identification for the 48 clusters with respective
kinematical analysis.  The results for each cluster are discussed in
the appendix. Then, in \S~5, we attempt a classification of typical cluster
morphologies and present our general results and discussions regarding
the kinematical and dynamical effect of the substructures. In \S~6 we
draw our conclusions.

Throughout, all errors are at  the 68\% confidence level (hereafter c.l.),
while the Hubble constant is 100 $h^{-1} \, Mpc^{-1} \,$ \ks.

\section{The Data Sample}

We apply our procedure to a set of 48 galaxy clusters, whose data are
taken mainly from literature sources and also from the recent {\it ESO
Nearby Abell Clusters Survey (ENACS)}\footnote{Concise information
on the ENAC Survey is available on the WWW at
http://www.eso.org/educnpubrelns/pr-05-96.html }, described in Katgert
et al. (1996). The clusters considered are Abell clusters except for
the poor cluster MKW3S, which belongs to the cluster field of A2063.

Only well-sampled clusters with a good level of completeness in
magnitude are suitable for detecting substructures.  In fact,
cluster regions which are oversampled with respect to the rest of the
cluster could produce artifacts which are not real substructures.
Also a galaxy sample randomly extracted from a magnitude-complete sample
is adequate to the study of substructures, although in this case there
is an obvious loss of information.

Here we considered only galaxy samples which nominally have the above
characteristics.  If necessary, we extracted from the whole data
sample a magnitude-complete one, according to the information given by
the authors.  When more than one redshift source is used, we checked
that a certain level of completeness is still conserved.  Out of the
48 clusters, ten include some data from ENACS in order to improve the
completeness in velocities (see Table~1); of these, A151 and A3128
have almost exclusively ENACS data.  We accept clusters with a minimum
level of magnitude completeness of 80\%.  For some clusters, for which
we do not have full information on magnitude, the completeness level
is considered acceptable by the authors.  In four cases (A539, A1060,
A2670 and A3526), we considered also an alternative initial sample,
indicated by an asterisk in Table~1, with a lower completeness level
or a smaller extension.  These alternative samples are considered less
useful for structure analysis and are used only to investigate or
confirm particular effects.  In all cases we refer to the authors for
the characteristics of completeness limits.

To fulfill   the   completeness requirements,  information   is  given
throughout the whole field down to  a limiting magnitude, and  therefore
the foreground and background objects are systematically included.\\
\end{multicols}

\renewcommand{\arraystretch}{1.0}
\renewcommand{\tabcolsep}{3.5mm}
\begin{center}
\vspace{-3mm}
TABLE 1\\
\vspace{2mm}
{\sc The Data Sample\\}
\footnotesize
\vspace{2mm}
\begin{tabular}{lrccccc}
\hline \hline
\multicolumn{1}{c}{Cluster Name}       & \multicolumn{1}{c}{$N_{field}$}                & 
\multicolumn{1}{c}{RS type}            & \multicolumn{1}{c}{Velocities Ref.}    & 
\multicolumn{1}{c}{Magnitudes Ref.}    & \multicolumn{1}{c}{$T_X$ Ref.}         &
\multicolumn{1}{c}{X-ray centers Ref.}  
\\
\multicolumn{1}{c}{(1)}       & \multicolumn{1}{c}{(2)}                & 
\multicolumn{1}{c}{(3)}       & \multicolumn{1}{c}{(4)}    & 
\multicolumn{1}{c}{(5)}       & \multicolumn{1}{c}{(6)}         &
\multicolumn{1}{c}{(7)}\\
\hline
A0085...................&     185&       cD&           3,38&  \nodata&12 &20\\
A0119...................&     139&        C&          21,23&    21,23&12 & 1\\
A0151...................&     142&       cD&          21,49&    16,21&18\tablenotemark{*}&18\\
A0193...................&      65&       cD&             31&  \nodata&12 &19\\
A0194...................&     200&        L&              9&        9&34 & 7\\
A0262...................&      88&        C&       25,28,42&  \nodata&12 &20\\
A0399...................&     227&       cD&             31&  \nodata&12 & 1\\
A0401...................&     227&       cD&             31&  \nodata&12 & 1\\
A0426...................&     128&        L&             35&    35,63&12 &20\\
A0539 (A0539*)...&189(153)&        F&             45&    43,64&12 & 1\\
A0548...................&     401&      (F)&          17,21&    17,21&13 &13\\
A0754...................&      89&       cD&             17&       17&12 &63\\
A1060 (A1060*)...&101(144)&        C&          52,53&    52,53&33 &20\\
A1146...................&      84&       cD&             58&  \nodata&57\tablenotemark{*}&46\\
A1185...................&      77&        C&              3&  \nodata&12 & 4\\
A1367...................&      68&        F&    15,24,26,59&       64&12 &29\\
A1631...................&      90&        C&             17&       17&18\tablenotemark{*}&18\\
A1644...................&     102&       cD&             17&       17&12 &20\\
A1736...................&     104&        I&             17&       17&12 & 6\\
A1795...................&      98&       cD&             31&  \nodata&12 &19\\
A1809...................&      69&       cD&          21,31&       21&18\tablenotemark{*}&18\\
A1983...................&     100&        F&             17&       17&18\tablenotemark{*}&18\\
A2052...................&      60&       cD&       21,38,50&    21,64&12 & 1\\
A2063...................&     141&       cD&           3,31&       16&12 & 4\\
A2107...................&      75&       cD&             44&  \nodata&12 &41\\
A2124...................&      67&       cD&             31&  \nodata&18\tablenotemark{*}&51\\
A2151...................&     106&        F&             17&       17&12 &51\\
A2197...................&      89&        L&             27&       64&30\tablenotemark{*}&51\\
A2199...................&      89&       cD&             27&       17&12 &51\\
A2634...................&     403&       cD&5,8,32,48,54,62&5,8,32,62&12 &53\\
A2666...................&     403&       cD&5,8,32,48,54,62&5,8,32,62&30\tablenotemark{*}&29\\
A2670 (A2670*)...&122( 88)&       cD&             55&       55&12 & 4\\
A2717...................&      81&cD\tablenotemark{b}&          10,21&    11,21&18\tablenotemark{*}&18\\
A2721...................&     104&       cD&          10,58&       11&18\tablenotemark{*}&18\\
A2877...................&     110&        C&             38&       16&12 &29\\
A3128...................&     222&        C&          10,21&    11,21&18\tablenotemark{*}&18\\
A3266...................&     172&       cD&             58&  \nodata&12 &29\\
A3376...................&      84&        L&             17&       17&18\tablenotemark{*}&18\\
A3391...................&     284&cD\tablenotemark{c}&             58&  \nodata&12 &29\\
A3395...................&     284&        F&             58&  \nodata&12 &29\\
A3526 (A3526*)...&123(105)&        F&          14,36&    14,36&12 &20\\
A3558...................&     551&cD\tablenotemark{b}&  2,21,39,52,58&  2,21,39&12 & 6\\
A3562...................&     551&  \nodata&  2,21,39,52,58&  2,21,39&12 &19\\
A3667...................&     177&L\tablenotemark{a}&          21,56&    21,56&12 &47\\
A3716...................&     106&        F&          10,17&       17&18\tablenotemark{*}&18\\
A3888...................&      98&        C&             58&  \nodata&60 &29\\
A4038...................&      51&F:(B)\tablenotemark{a}&          22,37&       22&12 &40\\
MKW3S...............&     141&       cD&           3,31&       16&61 & 4\\
\hline
\end{tabular}

\end{center}
{\footnotesize\parindent=3mm
REFERENCES. -- RS types:
(a) Bahcall 1977;
(b) Merrifield \& Kent 1991;
(c) Teague, Carter, \& Gray 1990.
Other references:
(1) Abramopoulos \& Ku 1983;
(2) Bardelli et al. 1994;
(3) Beers et al. 1991;
(4) Beers \& Tonry 1985; 
(5) Bothun \& Schombert 1988;
(6) Breen et al. 1994;
(7) Burns et al. 1994;
(8) Butcher  \& Olmer 1985;
(9) Chapman, Geller, \& Huchra 1988;
(10) Coless \& Hewett 1987;
(11) Coless 1989;
(12) David et al. 1993;
(13) Davis et al. 1995;
(14) Dickens, Currie, \& Lucey 1986;
(15) Dickens \& Moss 1976;
(16) Dressler 1980;
(17) Dressler \& Shectman 1988a;
(18) Ebeling et al. 1996;
(19) Edge \& Stewart 1991a;
(20) Elvis et al. 1992;
(21) ENACS;
(22) Ettori, Guzzo, \& Tarenghi 1995;
(23) Fabricant et al. 1993;
(24) Gavazzi 1987;
(25) Giovannelli, Haynes, \& Chincarini 1982;
(26) Gregory \& Thompson 1978;
(27) Gregory \& Thompson 1984;
(28) Gregory, Thompson, \& Tifft 1981;
(29) HEASARC Archive (NASA/Goddard) ;
(30) Henriksen 1992;
(31) Hill \& Oegerle 1993;
(32) Hintzen 1980;
(33) Ikebe et al. 1994;
(34) Jones \& Forman 1984;
(35) Kent \& Sargent 1983;
(36) Lauberts \& Valentjin 1989;
(37) Lucey \& Carter, 1988;
(38) Malumuth et al.  1992;
(39) Metcalfe, Godwin, \& Spenser 1987;
(40) McHardy et al. 1981;
(41) McMillan et al. 1989;
(42) Moss \& Dickens 1977;
(43) Nilson 1973;
(44) Oegerle \& Hill 1992; 
(45) Ostriker et al.  1988;
(46) Pierre et al. 1994;
(47) Piro \& Fusco-Femiano 1988;
(48) Pinkney, Rhee, \& Burns 1993; 
(49) Proust et al.  1992;
(50) Quintana et al.  1985;
(51) Rhee \& Latour 1991;
(52) Richter 1987;
(53) Richter 1989;
(54) Scodeggio et al.  1995;
(55) Sharples, Ellis, \& Gray 1988;
(56) Sodr\`{e} et al.  1992;
(57) Soltan \& Henry 1983;
(58) Teague, Carter, \& Gray 1990;
(59) Tifft 1978;
(60) White \& Fabian 1995;
(61) Yamashita 1992;
(62) Zabludoff et al. 1993;
(63) Zabludoff \& Zaritsky 1995;
(64) Zwicky et al. 1961-1968.}

\vspace{-3mm}
\normalsize

\begin{multicols}{2}
In the   same spirit, close  clusters  belonging to  a  given area are
considered as a single field :  the procedure we  apply is supposed to
retrieve  individual   clusters,  but with  objective centers   and
membership which may  differ from  traditional  ones.   Hence, we did
consider in the same  field  six associations of clusters:  A399-A401,
A2063-MKW3S,  A2197-A2199,  A2634-A2666,  A3391-A3395, and   the whole
region of the Shapley concentration (containing clusters 
A3558 and A3562).

Here we  discuss only  cluster  fields which appear to have at  least 50
galaxies  with available  redshift in the  main  peak of  the velocity
distribution (see next section).
Relevant entries for the clusters considered are shown in Table~1.  In
Col.~(1) we list the cluster names; in Col.~(2) the number of galaxies
with measured redshift in each cluster field; in Col.~(3) the
Rood-Sastry type given by Struble \& Rood (1987) for Northern
clusters and mainly by Struble \& Ftaclas (1994) for Southern
clusters; in Cols.~(4) and (5) the redshift and magnitude references,
respectively; in Cols.~(6) and (7) the references for the adopted
X-ray temperatures, hereafter $T$, and X-ray centers, respectively.

The X-ray temperatures indicated by an asterisk in Table~1 are rough
estimates coming from the X-ray luminosity, hereafter $L$, by means of
the $kT-L$ relation for the respective luminosity band.  In
particular, we used the relation by Edge \& Stewart (1991a) for A1146.
For A2197 and A2666 we used the relation $kT=10^{-7}\cdot
L(0.5-3KeV)^{0.17}$, which we obtained for the 28 clusters in common
between the samples of David et al. (1993) and Henrikensen (1992) by
means of a direct linear regression.

\section{The Method}

The purpose of this work is to point out 
 any physical structure which is
present within the analyzed  sample.  That implies, firstly, identifying
the cluster itself within its own environment, and  then detecting the
presence of any subsystem lying in the region.

Due to the complex gravitational phenomena occurring in the region of
a cluster, every physical structure is indeed  characterized by
correlations in space and velocity.  Therefore, no individual
structure identification can be performed without taking into account
both kinds of information simultaneously.  A direct three-dimensional
analysis is not suitable, since redshifts inside a cluster are not pure
distance parameters, owing to individual galaxy motions within the
cluster.  But redshifts still contain the traces of the dynamical
processes that form or dissolve the systems.  Depending on
their nature, some processes  will create local departures in
the observables, the most usual ones being the mean and the standard
deviation of the measured radial velocities (hereafter referred to
respectively as the mean velocity and the dispersion $\sigma$).
Hence, the analysis of local kinematics  appears a very suitable tool
to study physical processes occurring in galaxy clusters. 

\subsection{ Investigation of local kinematics.}

In this paper we call {\em  structure} any galaxy association which may be
physically connected through gravitational processes.  As stated in a
previous work (E94), an exploitable sign of the presence of these
structures within their environment is a local departure in their mean
velocities and dispersions.  To emphasize the dynamical processes,
which are present within a cluster of galaxies, we apply to each galaxy
the weight term $\rho$ which is a measure of
that galaxy's kinematics.
Such weights do work as local kinematical estimators since the
structure quantities are computed within a limited area of radius
$R_s$ around each galaxy.  The size of that area is related to the scale
of the exploration (see below) and so, as previously stated, the
crucial aspect of this analysis is the use of a multi-scale approach.

In  the   following,  $\sigma$  is     the   velocity  dispersion   and
$\overline{v}$ is the  mean velocity of  the $n$ galaxies found in the
area considered; the label {\em  loc} refers to the local area of radius
$R_s$, while the label {\em  main} refers to the whole field.

The first estimator  looks  for the  local  departure of the  velocity
dispersion~:

\begin{equation}
\rho_S = n       \times (\sigma_{main} / \sigma_{loc} )^2. 
\end{equation}

\noindent 
In this way, low values of local dispersions will produce high values 
of $\rho_S$.

The second estimator searches for local departures of the mean velocity :

\begin{equation} 
\rho_V = n  \times 
\left( \frac
{\overline{v}_{loc} - \overline{v}_{main} }{\sigma_{main} }\right)^2. 
\end{equation}

\noindent 
Hence, prominent departures in the local mean velocity (e.g. by over 
a $\sigma_{main}$) will produce high values of $\rho_V$.

We normalize  the weight  terms  $\rho$ to their own mean
values within  the whole cluster:

\begin{equation} 
\delta = \rho / {\overline \rho}. 
\end{equation}

\noindent 
In this way the value of $\delta$ does not
depend on the cluster analyzed and its mean value is equal to one.  Hence,
values departing from unity directly identify the expected local
effects.

Now we form a third estimator by computing the quadratic sum of the previous 
two~:

\begin{equation} 
\delta_P = \sqrt{ \delta_S^2  +  \delta_V^2 }.  
\end{equation}

Finally we include a fourth estimator which is a local version of 
the Dressler parameter (Dressler \& Shectman 1988b, hereafter D88)~: 

\begin{equation} \delta_D =\frac{n_{loc}}{ \sigma_{main} } \times 
\sqrt{ (\overline{v}_{loc} - \overline{v}_{main})^2  +  
   (\sigma_{loc} - \sigma_{main})^2 }.  
\end{equation}

The estimator $\delta_P$ differs from $\delta_D$ since the
normalization on $\rho_S$ and $\rho_V$ is done in eq.~4 before summing
rather than after summing (as in eq.~5 for $\delta_D$). In this way,
the estimator $\delta_P$ takes into account the departures
in velocity and dispersion separately, while these two effects are confused in
$\delta_D$.  It must be noticed that E94 did use the true Dressler
parameter for nearly the same purpose.  The main difference compared to
the above $\delta_D$ is the restriction to a limited area around the
galaxy considered, while the Dressler parameter systematically
includes 11 neighbours in the computation, irrespective of the
distance.  The present work includes some clusters already considered
in E94, in a few cases with the same dataset; using the kinematical
weights mentioned above, we now expect  to detect some new structures,
e.g. those exhibiting discrepancies in the local dispersion or in the
local mean velocity, whilst some previously detected structures will
no longer appear significant, according to the more local definition
of the present analysis.

The confidence level of the weight values is derived from the
statistics computed on the set of values obtained throughout the whole
cluster.  High values correspond to prominent events, which occur around
the galaxy considered.  In this work we adopt the classical 3
s.d. threshold above the mean, which here refers to the statistics on
the weight values. 
 In order to obtain a more reliable estimate of the confidence level,
we compute the distribution of weights for the whole 
sets of replicas (see below).
In this way, for a given scale of analysis, we obtain a list
of galaxies which are presumed to identify a structure.

\subsection{Investigation of the subclustering processes.}

Once the technique of  individual weighting is applied, the  spatial
clustering of the galaxies needs   to be quantified and estimated   in
terms  of statistical validity.  Therefore, we   choose to perform  the
wavelet   transform,  which   is  particularly well    suited  for such
purposes.  The   ability of  galaxies to   form structures is measured
objectively  with the wavelet  coefficients in a multi-scale way. 

The starting  dataset    for  wavelet  analysis  is a    bidimensional
distribution of  {\em  weighted  galaxies},  viz. a distribution   of Dirac
functions normalized to the weights  $\delta_i$. The analysis consists
then in performing the transform by a wavelet function.

The details of analyzing a spatial distribution of galaxies with the
wavelet transform have been extensively described in a series of
previous papers (Escalera \& MacGillivray 1995 and references
therein).  In this work we use the so-called Mexican Hat wavelet,
which makes the transform at a given scale insensitive to the presence
of a gradient at a much different scale (see also E94).

Therefore, the full-scale analysis is not sensitive to the presence of
small-scale structures and leads to a definition of the main system.

The  main  procedure   of   our  analysis  consists in    investigating
simultaneously   the   local kinematics   (weights) and    the spatial
clustering (wavelets). The two  techniques  are fully consistent with  each
other, since   they are both     objective (no free  parameters and   no
preliminary assumptions  are  needed)   and  they both consist    in  a
multi-scale analysis.  The point is to use the same range of explored
areas   for  the weight terms   and  for  the  wavelet transform. When
investigating a spatial distribution at  a given scale $s$, the Mexican
Hat  explores areas roughly  extended  for 4$s$.   Thus,  consistency
with the  weighting procedure is obtained by  simply exploring an area
of  the same radius $R_s =  4 s$ at  the time of  computing the weight
values.

In practice  we adopt  a series  of  three successive  scales~:  $s$ =
0.03125, 0.0625, and 0.09375, leading to explored areas of 
 radius $R_s$ equal to  0.125,   0.25, and  0.375, respectively, in units of 
the maximum radius of the field analyzed.   Such a limited series appears
sufficient  to retrieve  any  substructure  present   in the
sample.  The only requirement for detecting  conveniently a structure of a
given size  is to use  an immediate upper and  lower scale.  It is
not the aim of this work to detect the small pairs  or triplets; thus the
lower limit of 0.03125 does fulfill the purpose.

The membership of a given structure is the set of galaxies, within the
explored area, selected by the weighting technique, i.e. galaxies
which have significant weights ($| \delta - \overline \delta | \ge 3
\sigma_{\delta}$, see \S~3.1) and confirmed by wavelet analysis.
Then, the estimate of structure size is determined on the identified
members by computing the projected radius.  Thus, it is possible that we
may retrieve some structures which are smaller than the smallest wavelet scale
we use.  Throughout the present work, the wavelet scales we use lead
to resulting structures with dimensions of about 1.5 \h (median value)
for the main cluster, down to about 0.2 \h for the smallest
substructures detected.

We preferred to use a relative array of scale sizes rather than a
fixed one because we study clusters of different intrinsic sizes for
which a fixed scale could have a different physical
interpretation. For instance, 0.5 \h can be the measure for the global
size of a poor cluster or the measure for a clump in a bimodal
cluster.  In our procedure the first step generally gives 
the cluster immediately, independently of its dimensions. Wavelet analysis
does not require the use of a scale rigorously equal to the size of the
structure, but it is only necessary to approach this value by close wavelet
scales. Hence, the use of a fixed array of scale sizes rather than a
relative one should not have strong repurcussions on the determination
of structure sizes if a similar range of sizes is examined.  In
particular, our evaluations of structure sizes depend on the values of
kinematical weights and so they are not strictly linked to the choice
of the array of wavelet scales.

The statistical significances are simply derived by comparing the
wavelet coefficients obtained in the real field with those produced in a
series of $N$ replicas (see e.g. Escalera \& Mazure 1992).  The replicas are
obtained by drawing independently the positions $X_i$ and $Y_i$ from
the $X$ and $Y$ distributions of the sample studied and then by randomly
reassigning the velocities.  These replicas contain all the phenomena
that can produce random associations of galaxies. By selecting the
groups which do not appear in the replicas we simply separate the
underlying physical processes from the random ones.  Thus we finally
arrive at the probability that the observed structure is not due to a
randomly associating process or to projection effects.\\

\end{multicols}
\includegraphics{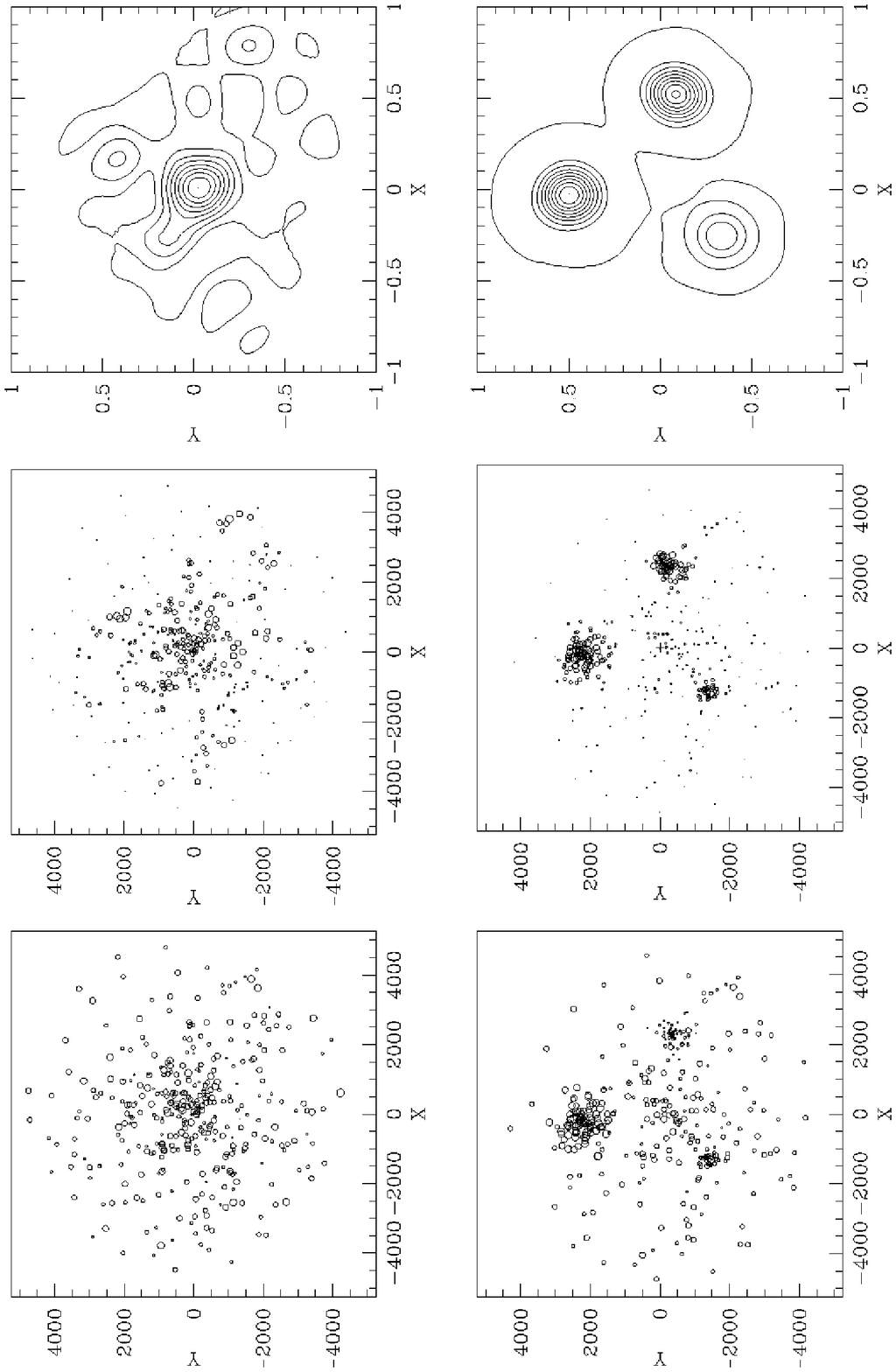}
$\ \ \ \ \ \ $\\
\vspace{11.4cm}
$\ \ \ $\\
{\small\parindent=3.5mm
{\sc Fig.}~1.---
Illustration on a toy-model.  A simulated
regular cluster (Figures on the top) compared to a perturbed cluster
(Figures on the bottom).  Figures at the left show the distribution of
galaxies: the symbol size increases towards low values of velocity.
Substructures in perturbed cluster are thus clearly visible.  Figures
at center show the distributions of the weighted galaxies: the symbol
size now increases towards high values of the weight $\delta_D$.  In
the case of the perturbed cluster, the selection of the significant
value locates departures in the local kinematics.  On the contrary, in
the regular cluster, none of the observed $\delta$ values is
significant.  Figures at the right show the wavelet images of the
above weighted bidimensional maps, i.e. the isophotes of the wavelet
coefficients.  As above, the perturbed cluster shows significant
features (compared to random simulations) which do correspond to the
input substructures, while the regular cluster does not contain any
significant features. 
}
\vspace{5mm}
\begin{multicols}{2}

  We point out that an appreciable improvement in the analysis comes
from the fact that departures in mean velocity and in dispersion are
investigated separately, by means of specific weight terms.

For each  cluster we  consider 4 weight   terms at 3 different  scales,
obtaining  in  that way  a  series of 12  maps.  We  retain a 
structure if it appears significant in at least one of these maps.

The  main results are  the  structure positions derived by the
location  of the local maximum  of the  wavelet coefficients, the full
membership given by the list of galaxies  responsible for the observed
local departure in  kinematics, and the significance level, which is
the probability for the observed structure to be reproduced within the
random replicas of  the analyzed data.  The membership  identification
makes possible a dynamical analysis of the structure.

\subsection{Illustration on a toy model.}

We include here an example of a practical application of the whole
detection analysis to a toy model, with the only purpose of
emphasizing the possibilities and the limits of our detection
procedure.  For a full illustration of our procedure, where we
consider many alternative toy models by varying the positions,
extents, and dynamics of the input substructures, we refer to Escalera
\& MacGillivray (1995) and references therein.

The simulations we use here consist in a regular cluster compared to a
perturbed cluster.  The regular cluster has a smooth symmetric density
profile (viz. the so-called King profile) and a Gaussian velocity
distribution.  The second cluster, similar to the regular one in
extension, population, and global kinematics, consists in a main fully
regular structure (M) perturbed by a loose triplet (D) and by three
regular substructures (A,B,C) which are distinct from each other in
population and extension and are inserted within the limits of M.  The
subgroups depart from M in terms of mean velocity and/or
dispersion (see Table~2 and Figure~1).\\

\hspace{-4mm}
\begin{minipage}{9cm}
\renewcommand{\arraystretch}{1.2}
\renewcommand{\tabcolsep}{3.mm}
\begin{center}
\vspace{-3mm}
TABLE 2\\
\vspace{2mm}
{\sc Detection in a toy model\\}
\footnotesize
\vspace{2mm}
\scriptsize
\begin{tabular}{crrrrr}
\hline
\hline
\colhead{Structure}
&\colhead{$N_{gal}$}
&\colhead{X\phm{00000}Y}
&\colhead{$\overline{v}$}
&\colhead{$\sigma$}
&\colhead{$P_{SL}$}\\
\hline
\multicolumn{6}{c}{INPUT}\\
\hline
total&   353 &        0.0\phn\phn\phn\phs 0.0 & 14944   &   1759 &\\
M    &   200 &$-$\phn16.2\phn   $-$\phn31.7 & 15070   &   1059 &\\
A    &    75 &$-$\phn12.2\phn     \phs493.0 & 12572   &    419 &\\
B    &    50 &      508.3\phn$-$\phn\phn9.0 & 17627   &    728 &\\
C    &    25 &   $-$214.3\phn      $-$208.5 & 15956   &     76 &\\
D    &     3 &      203.0\phn     \phs403.0 & 12003   &\nodata &\\
\hline
\multicolumn{6}{c}{OUTPUT}\\
\hline
M & 196 &$-$\phn13.5\phn   $-$\phn27.1 & 15040   &   1066 &\nodata\\
A &  75 &$-$\phn12.0\phn     \phs490.8 & 12578   &    434 &   0.00\\
B &  51 &      506.8\phn$-$\phn\phn7.9 & 17602   &    754 &   0.00\\
C &  28 &   $-$214.2\phn      $-$209.6 & 15952   &     92 &   0.00\\
D &   3 &      203.0\phn     \phs403.0 & 12003   &\nodata &   0.01\\
\hline
\end{tabular}

\end{center}
\vspace{3mm}
\end{minipage}
\normalsize

As one  can see in Table~2, A  is almost completely detected, with one
true object missing and one   false object (interloper)  included, and
its dynamics is  accurately obtained. 
 B  is fully detected with  the
addition of a single false object which does not significantly perturb
the main dynamics of the structure. C is also fully detected, with the
addition  of 3    false   objects,  so its dispersion     is  slightly
overestimated,  though  within the error  bars, and its  mean velocity
remains acceptable since - by definition - the contaminating objects are
related  to   the real dynamics   of the  subgroup. Finally,  the close
triplet   D is detected with  no   contamination.  The resulting  main
cluster M is  obtained by subtraction  of  the detected substructures,
and  consequently  appears very close   to the input  data:  only 5
objects are missing (wrongly  attributed to the substructures) and one
is added.
Then, as expected, no clump is significantly detected in the regular
cluster.

In Table~2 we list the structure positions, i.e. center coordinates in
arbitrary units with a maximum error of 7.8 units; the membership,
which leads to the computation of the mean velocity $\overline{v}$ and
dispersion $\sigma$; the significance level, i.e. the percentage of
similar structures found in the simulations.  Through the whole paper
we used a series of 1000 simulations to compute the significance
levels.

When showing the detection power of a method, it is also important to
keep in mind its limits. We stress that the structures are
detected if they depart significantly from the environment (cluster
from field, or substructure from cluster), i.e.  if
$\overline{v}_{loc}$ departs from $\overline{v}_{main}$ by more than
$0.3 \times \sigma_{main}$, and/or $\sigma_{loc}$ is smaller than
$0.8 \times \sigma_{main}$.  Moreover, the membership is retrieved
with an error of about 10\%, i.e.  one object out of 10 can be missed
and/or replaced by a contaminating object.

 Obviously, the position of the structure within the cluster
and its relative extension affect the efficiency of the detection.  In
practice, structures that do not obey at least one of the above two
criteria will be missed; e.g. structures with low departures of
$\overline{v}$ are not significant if they are not well separated in
the map.  They just resemble projected random fluctuations of the
3-$D$ distribution.
On the contrary, structures that fulfill the two above conditions
can  be easily detected   whatever their  relative  population and
location within the cluster field may be.

\subsection{Procedure on real data}

The application  to real data of the procedure described above requires a 
preliminary identification of the true cluster in  the field.
The full system identification  therefore consists in a series of three 
successive stages arranged as follows.\\

{\em 1. Main Peak.  } Several methods have been proposed in the
literature for identifying coherent physical systems in redshift
surveys (see, e.g., Mazure et al. 1996).  In this preliminary stage,
however, we do not want to prematurely break up physical systems.  We
only want to identify obvious subsystems (fore- and background
groups), keeping the dominant system intact for further
three-dimensional analysis. In order to avoid unnecessary
sophistication, we used the Poissonian Gap method, which is a simple
and stable method for defining systems.  The gap is the separation
between adjacent galaxies in the velocity distribution.  For each
cluster field, gaps greater than the median value generally correspond
to unrelated systems, as recently demonstrated by Katgert \&
al. (1996).  Such conclusions do not depend on noise effects as long
as the samples are complete enough and do contain a reasonable
population (namely $\geq$ 30 galaxies).  Both conditions are fulfilled
in our dataset.

When more than one system is found in the field, for further analysis
we retain only those with at least 50 galaxies.\\

{\em  2. Main System. } The present stage introduces the three-dimensional
analysis and consists in exploring the samples at a large scale ($s$ =
0.125). In this way, the considered area approaches the whole field ($R_s$
= 0.5); thus the local kinematics is close to that of the whole field.
Therefore, at this stage, we spatially identify the true cluster
within the selected main peak.  Since in any case we combine 
positions and redshift information, the remaining
background/foreground galaxies are also identified and removed.
These  galaxies  initially belong to the main peak
of the velocity distribution, but are indeed outside the spatial
limits of the detected structure.
In the case of bimodal clusters and of two distinct populations, which
overlap in velocity, we can clearly separate them by combining galaxy
positions with kinematic estimators.
At this stage unconnected subsystems are listed and removed for 
further analysis.\\

{\em  3. Multi-scale analysis. } We calculate the weights (described in
\S~3.1) for each galaxy by taking into consideration a surrounding area,
whose size corresponds to the chosen scale.  As previously stated, the
multi-scale analysis consists in taking a series of decreasing scales.

\end{multicols}
\includegraphics{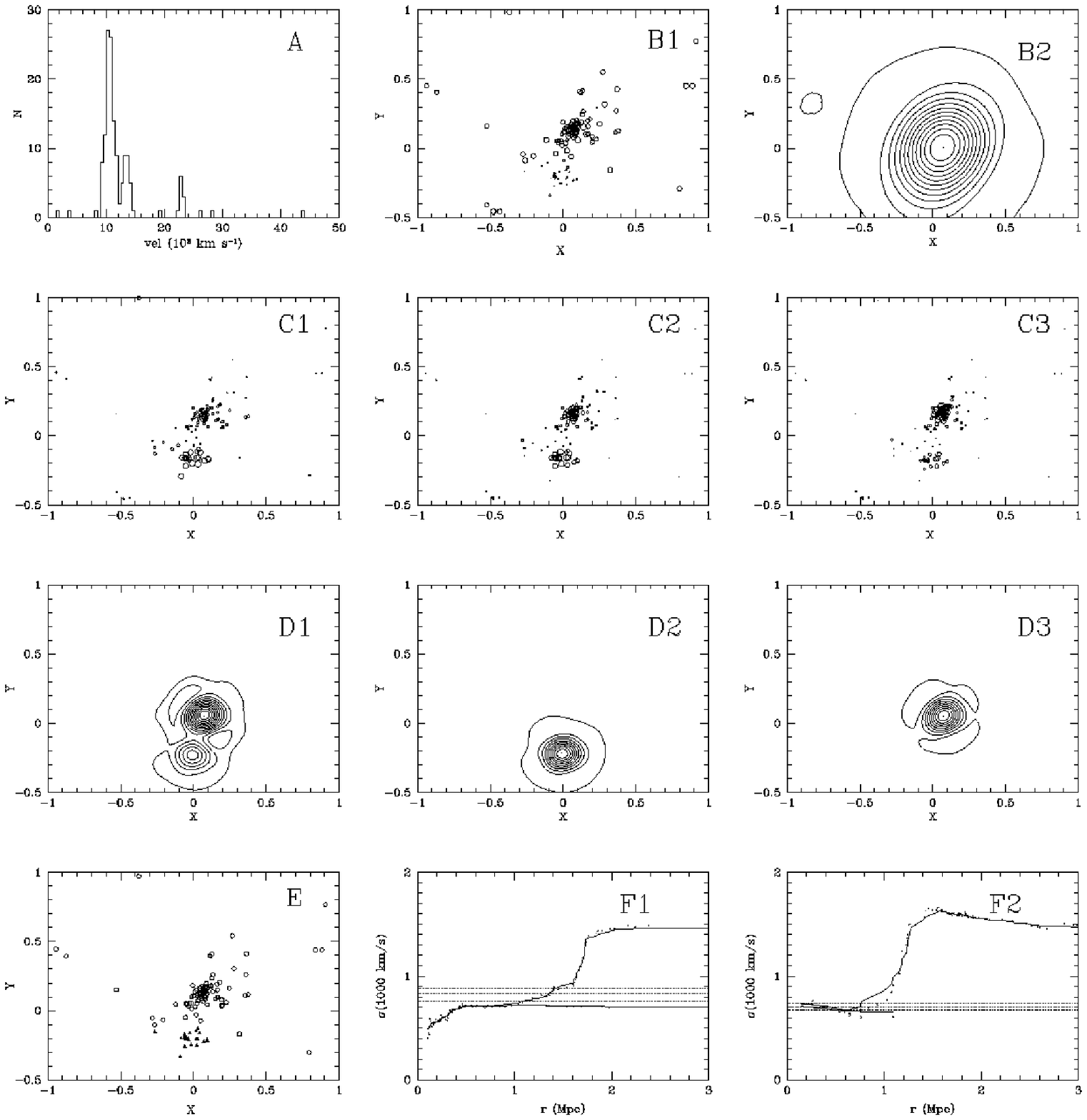}
$\ \ \ \ \ \ $\\
\vspace{18.7cm}
$\ \ \ $\\
\vspace{1mm}
{\small\parindent=3.5mm
{\sc Fig.}~2.---
A real example: A2063-MKW3S field. (A) Selection
by gapper procedure of the main peak in the redshift distribution
(roughly from 9000 to 15000 \ks); (B) Spatial distribution of galaxies
in the main peak (symbol sizes increases towards low redshifts) and
corresponding isopleths of the wavelet coefficients, as a result of
the large scale bi-dimensional analysis; (C) Substructure
identification by applying the weight terms $\delta_D$, $\delta_V$,
and $\delta_S$ respectively, with large symbols corresponding to large
departure in the local kinematics; (D) Isopleths of the wavelet
coefficients obtained through analysis of the {\em weighted } cluster
as in C1, C2, and C3 respectively. Both structures are present in the
$\delta_D$ image (D1), while the $\delta_V$ image (D2) shows only the
background structure, whose departure in the local mean velocity is
high. The $\delta_S$ image (D3) shows the main concentration; there
are, in fact, no strong deviations from the local dispersion in this
sample, so only spatial clustering is retrieved; (E) The membership of
the two structures indicated by two different symbols; (F) Comparison
between the velocity dispersion profiles of the initial main sample
and the two selected structures, main concentration (F1) and
background structure (F2), respectively. In computing the profiles the
respective X-ray centers are used. The corresponding X-ray
temperatures with their error bands are plotted for comparison.}
\vspace{5mm}
\begin{multicols}{2}

The above procedure can be summarized with the following symbols:

\begin{itemize}
\item {\it field}, the initial sample (whole cluster field).
\item {\it MP}, the main peak, which results from the Poissonian gap method.
\item {\it US}, or {\it US1, US2, etc.}, unrelated structures, are
coherent systems unrelated to the cluster and are identified from the
bi-dimensional analysis at the largest scale. An US structure is
considered as a secondary main system if its population is at least
$\sim 25\%$ of the primary main system (see below MS1, MS2).
\item {\it MS}, the main system, 
which generally corresponds to the {\em identified
cluster}.
\item {\it MS1, MS2}, two comparable main structures, e.g.  two
clusters in the same field or individual lobes in the case of a
bimodal cluster.
\item {\it S} or {\it S1, S2}, etc., the successive substructures,
outputs of the multi-scale structural analysis.
\item {\it C}, the core structures, i.e. structures detected in the
central cluster region, whose mean velocities do not significantly
differ from the respective cluster mean velocities (i.e.  the
difference is less than the velocity dispersion of the cluster
itself).

\end{itemize}

Sometimes we needed to analyze the effect of removing a substructure 
from the parent structure. We refer to the
remaining galaxies by inserting a sign of subtraction between the
symbols of structures, e.g. MS-S1-S2 if  substructures
S1 and S2 are removed from the MS structure. \\

For each cluster we have to examine twelve figures (four weights at
three different scales).  We illustrate the procedure described above
by giving the complete set of figures for the field of A2063-MKW3S
(see Figure 2). It consists of a main regular cluster with a poor
background cluster which is only $\sim$ 3000 \ks away and thus gives a
clear representation of the way the method works.

\section{The Detected Structures}

The results of our structural analysis (\S~3) of galaxy clusters are
presented in Table~3, which also contains the basic kinematical
properties of the detected structures.  The structures mentioned have
a confidence level $\geq$ 99.5\%, i.e. less than 5 chances in 1000 of
being due to a random configuration. In some few cases, however, we do
include results for less significant structures which appear to be of
some particular interest (as specified in Table~3). In principle, each
structure corresponds to a physical structure. Artifacts and field
contamination are not touched in the discussion.

We applied homogeneous procedures to the study of the detected
structures. In order to determine the center of structures, we used
the two-dimensional application of the adaptive kernel method (e.g
Pisani 1993; Girardi et al. 1996 and reference therein).  

Then, the projected radius $R$ of the structures is determined
as the maximum projected distance from the center  for all the galaxies
belonging to the structure.  

We used robust mean and dispersion estimates computed by using the
ROSTAT routines by Beers, Flynn \& Gebhardt (1990).  As an estimator
of the Gaussianity of velocity distributions, we adopted the
probability $P_W$ associated with the W-test (Shapiro \& Wilk 1965).
Remarkably, non-Gaussian velocity distributions could be due to the
presence of substructures, but also to the presence of velocity
anisotropies (Merritt 1987); thus the absence of Gaussianity is only a
{\em sign } of possible substructures.

Hence, for all the detected structures, Table~3 gives the following
entries: the field name and their nature, indicated by symbols as
described in \S~3.4; the number of involved galaxies $N$; $\alpha$ and
$\delta$ coordinates of the galaxy density center, epoch 1950; the
overall projected radius $R$ (in \h); the mean velocity
$\overline{v}$; the $P_W$ probability and velocity dispersion $\sigma$
(in \ks) with the respective bootstrap errors; 
the name of the {\em  identified  cluster}: 
this identification is
particularly useful when the field contains more than one cluster.

The spatial distributions of galaxies of {\em identified clusters }, which show
internal structures (substructures and/or core structures), 
are displayed in Figure~3.

For each {\em  identified cluster}, and other interesting structures, in
Figure~4 we show the respective velocity dispersion profile, hereafter
VDP, which, at a given radius, is the averaged l.o.s. velocity
dispersion within this radius. 
The horizontal lines in Figure~4 show the values of the velocity
dispersion obtained from the temperatures (see Table~1) under
the condition of a perfect galaxy/gas energy equipartition, i.e.
with $\beta=\sigma^2/(kT/\mu m_p)=1$, where $\mu$ is the mean molecular
weight and $m_p$ the proton mass (see e.g. Sarazin 1986 and references
therein).  If both galaxies and gas are in dynamical equilibrium
within the cluster, one expects that the observed $\sigma$ will coincide
with that obtained from $T$.

  The square of l.o.s. velocity dispersion, as computed on the whole
cluster, is a third of the squared spatial velocity dispersion independently
of the presence of velocity anisotropies in galaxy orbits (e.g. The \&
White 1986; Merritt 1988). However, velocity anisotropies can strongly
influence the l.o.s.  velocity dispersion, as computed on the central
cluster region.  In particular, the presence of circular orbits in the
central cluster region, as expected in a relaxed cluster, produces a
VDP decreasing towards the cluster center (e.g. Sharples et
al. 1988). On the other hand, the presence of radial orbits in the
external region, as expected for a cluster with galaxy infall,
produces a VDP increasing towards the cluster center (e.g. Merritt
1987).  The VDPs of our {\em identified clusters} are generally flat
in the external regions. \\
\end{multicols}

\renewcommand{\arraystretch}{1.0}
\renewcommand{\tabcolsep}{3.5mm}
\begin{center}
\vspace{-3mm}
TABLE 3\\
\vspace{2mm}
{\sc Detected Structures\\}
\vspace{2mm}
\tiny
\begin{tabular}{llrrrrrrl}
\hline
\hline
\\
\multicolumn{2}{c}{Sample name}
&\colhead{$N$}
&\colhead{Center 1950 ($\alpha$,$\delta$)}
&\colhead{$R$}
&\colhead{$<V>$}
&\colhead{$P_W$}
&\colhead{$\sigma$}
&\colhead{Identified}
\\
\multicolumn{2}{c}{}
&\colhead{}
&\colhead{}
&\colhead{(Mpc)}
&\colhead{(Km/s)}
&\colhead{}
&\colhead{(Km/s)}
&\colhead{cluster}\\
\\
\hline
\\
A0085............&MP=MS&124&003904.9$-$093356 &   1.74&  16605& 0.1&1015\phn($-$\phn 72,\phs$+$\phn 83)&A85\\
                        &    S&  7&003906.5$-$093528 &   0.22&  13869&32.7& 352\phn($-$\phn 17,\phs$+$252)&\\
A0119............&MP=MS&123&005345.9$-$013125 &   1.26&  13258& 7.5& 769\phn($-$\phn 61,\phs$+$\phn 69)&\\
A0151............&   MP& 95&010621.2$-$154039 &   1.73&  15062&$<0.1$&1860\phn($-$133,\phs$+$108)&\\
                        &  MS1& 65&010621.4$-$154036 &   1.59&  15952&63.5& 708\phn($-$\phn55,\phs$+$\phn 69)&A151\\
                        &  MS2& 28&010625.5$-$161408 &   0.82&  12317&47.8& 391\phn($-$\phn 43,\phs$+$\phn 77)&\\
A0193............&MP=MS& 56&012228.2\phs082619&   0.89&  14559&26.0& 726\phn($-$\phn 61,\phs$+$\phn 78)&A193\\
                        &    S&  4&012236.2\phs082328&   0.04&  14648& 8.2& 171\phn($-$\phn 33,\phs$+$162)&\\
A0194............&   MP&156&012317.5$-$013644 &   4.55&   5348&$<0.1$& 579\phn($-$\phn 76,\phs$+$185)&\\
                        &  US1& 16&012053.1\phs012307&   3.85&   9683&22.7& 490\phn($-$\phn 52,\phs$+$\phn 90)&\\
                        &  US2& 15&012033.1$-$005137 &   4.94&   8381& 7.6& 619\phn($-$\phn 40,\phs$+$128)&\\
                        &   MS&121&012317.6$-$013640 &   4.55&   5354& 7.6& 426\phn($-$\phn 31,\phs$+$\phn 46)&A194\\
                        &   S1&  3&013129.7$-$011852 &\nodata&\nodata&\nodata&\nodata\phm{++000000,\phs()}&\\
                        &   S2&  7&011022.8$-$003321 &   0.42&   5390&91.1& 178\phn($-$\phn 23,\phs$+$\phn 59)&\\
                        &    C& 19&012319.2$-$013553 &   0.98&   5279&10.3& 249\phn($-$\phn 29,\phs$+$\phn 43)&\\
A0262............&   MP& 83&014951.1\phs355314&   5.77&   4904&12.9& 538\phn($-$\phn 37,\phs$+$\phn 54)&\\
                        &   MS& 48&014949.1\phs355310&   2.53&   4881&23.8& 528\phn($-$\phn 42,\phs$+$\phn 59)&A262\\
A0399$-$401....&     MP&214&025613.6\phs132434&   3.72&  21838&10.0&1201\phn($-$\phn 45,\phs$+$\phn 70)&\\
                        &  MS1& 86&025612.8\phs132433&   1.99&  21901&15.0&1012\phn($-$\phn 60,\phs$+$\phn 76)& A401\\
                        &  MS2& 79&025507.3\phs124755&   1.44&  21151& 1.4& 961\phn($-$\phn 55,\phs$+$\phn 71)& A399\\
A0426............&   MP&126&031621.8\phs412155&   2.65&   5243& 3.7&1239\phn($-$\phn 97,\phs$+$115)&\\
                        &   MS&122&031621.8\phs412155&   2.64&   5228&68.2&1138\phn($-$\phn 80,\phs$+$\phn 92)&A426\\
                        &    C& 37&031624.1\phs412200&   0.53&   5081&37.9&1154\phn($-$\phn 99,\phs$+$133)&\\
A0539............&   MP&177&051917.4\phs032341&  13.19&   7061&$<0.1$&2415\phn($-$\phn 99,\phs$+$\phn 59)&\\
                        &  MS1& 65&045617.1$-$003552 &   9.89&   4478&$<0.1$& 382\phn($-$\phn 53,\phs$+$\phn 96)&\\
                        &   C1& 25&045616.5$-$003541 &   2.73&   4428&59.4& 267\phn($-$\phn 27,\phs$+$\phn 40)&\\
                        &  MS2& 85&051357.3\phs062413&  13.72&   8619& 6.6& 449\phn($-$\phn 39,\phs$+$\phn 57)&\\
                        &   C2& 19&051733.3\phs063202&   2.60&   8797& 7.1& 227\phn($-$\phn 43,\phs$+$\phn 70)&A539\\
A0539*..........&     C& 36&051355.6\phs062334&   0.36&   8662&27.6& 985\phn($-$\phn 76,\phs$+$108)&\\
                        &   C1& 23&051354.0\phs062655&   0.36&   8112&11.1& 620\phn($-$\phn 85,\phs$+$\phn 89)&\\
                        &   C2& 13&051409.0\phs062758&   0.49&   9709&11.8& 402\phn($-$\phn 36,\phs$+$\phn 78)&\\
A0548............&   MP&341&054326.9$-$255739 &   2.61&  12416& 0.8& 842\phn($-$\phn 24,\phs$+$\phn 29)&\\
                        &  MS1&190&054326.8$-$255738 &   1.51&  12603& 0.2& 880\phn($-$\phn 31,\phs$+$\phn 40)&A548SW\\
                        &   S1& 12&054112.6$-$255734 &   0.16&  11645&68.0& 691\phn($-$\phn 83,\phs$+$136)&\\
                        &   S2& 43&054327.3$-$255726 &   0.56&  13047& 5.3& 657\phn($-$\phn 63,\phs$+$\phn 63)&\\
                        &   S3& 17&054233.7$-$263538 &   0.41&  11862&63.4& 303\phn($-$\phn 33,\phs$+$\phn 61)&\\
                        &   S4& 17&054226.8$-$260540 &   0.48&  12192&22.2& 643\phn($-$\phn 49,\phs$+$\phn 96)&\\
                        &  MS2&149&054636.5$-$253110 &   1.16&  12167&$<0.1$& 680\phn($-$\phn 26,\phs$+$\phn 30)&A548NE\\
A0754............&   MP& 89&090711.0$-$093108 &   2.64&  16257&$<0.1$& 817\phn($-$\phn 77,\phs$+$130)&\\
                        &   US&  8&090535.2$-$094736 &   0.72&  15889&10.0& 477\phn($-$\phn 89,\phs$+$385)&\\
                        &   MS& 38&090708.3$-$093049 &   1.42&  16428& 0.7& 495\phn($-$\phn 56,\phs$+$\phn 82)&\\
                        &  MS1& 22&090618.7$-$092155 &   0.76&  16218& 0.4& 409\phn($-$\phn 17,\phs$+$109)&A754NW\\
                        &  MS2& 16&090710.1$-$093019 &   0.88&  16717& 3.9& 531\phn($-$\phn 92,\phs$+$110)&A754SE\\
                        &    C&  8&090709.9$-$093023 &   0.90&  16950&12.9& 607\phn($-$\phn 61,\phs$+$118)&\\
A1060............&MP=MS& 94&103411.4$-$271436 &   2.27&   3752& 8.2& 634\phn($-$\phn 41,\phs$+$\phn 45)&A1060\\
                        &    S&  5&103217.8$-$281905 &   0.05&   3402& 0.7& 130\phn($-$\phn\phn 2,\phs$+$\phn 22)&\\
                        &    C& 14&103414.0$-$271456 &   0.27&   3881&13.6& 748\phn($-$\phn 74,\phs$+$117)&\\
A1060*..........& MP=MS&125&103411.6$-$271442 &   2.26&   3739&17.8& 633\phn($-$\phn 32,\phs$+$\phn 47)&\\
                        &    C& 40&103411.7$-$271428 &   0.79&   3690& 3.9& 780\phn($-$\phn 51,\phs$+$\phn 63)&\\
A1146............&MP=MS& 61&105850.1$-$222806 &   1.72&  42646&73.7&1028\phn($-$\phn 96,\phs$+$\phn 93)&A1146\\
A1185............&   MP& 69&110802.2\phs285803&   4.51&   9127& 1.8&1240\phn($-$\phn 90,\phs$+$123)&\\
                        &   MS& 55&110802.3\phs285803&   4.64&   9470& 3.2& 786\phn($-$\phn 54,\phs$+$\phn 54)&A1185\\
                        &    C& 23&110744.8\phs285746&   1.27&   9344&22.4& 567\phn($-$\phn 46,\phs$+$\phn 88)&\\
A1367............&MP=MS& 68&114137.4\phs200601&   0.98&   6432&40.0& 838\phn($-$\phn 68,\phs$+$\phn 81)&A1367\\
                        &    S&  6&114119.8\phs201415&   0.09&   6391&49.3& 330\phn($-$\phn 22,\phs$+$103)&\\
A1631............&MP=MS& 71&125020.6$-$150453 &   1.96&  13962&48.9& 703\phn($-$\phn 47,\phs$+$\phn 54)&A1631\\
                        &    C& 15&125019.7$-$150434 &   0.65&  13583&18.7& 530\phn($-$\phn 76,\phs$+$100)&\\
                        &   C1&  6&125019.5$-$150431 &   0.13&  13704&41.0& 310\phn($-$\phn 56,\phs$+$131)&\\
A1644............&   MP& 91&125444.4$-$170748 &   1.98&  14120&14.1& 927\phn($-$\phn 78,\phs$+$\phn 89)&\\
                        &   MS& 84&125446.2$-$170939 &   1.90&  14020&76.7& 763\phn($-$\phn 50,\phs$+$\phn 64)&A1644\\
A1736B..........&    MP& 63&132426.5$-$265337 &   2.14&  13812& 4.7& 976\phn($-$\phn 64,\phs$+$\phn 66)&\\
                        &   MS& 51&132428.3$-$265435 &   1.54&  13594& 2.6& 824\phn($-$\phn 47,\phs$+$\phn 65)&A1736B\\
A1795............&   MP& 85&134632.4\phs264905&   1.79&  18888& 0.1& 873\phn($-$\phn 75,\phs$+$121)&\\
                        &   MS& 83&134632.4\phs264905&   1.79&  18885&56.4& 828\phn($-$\phn 72,\phs$+$\phn 88)&A1795\\
                        &    C& 28&134629.9\phs264825&   0.41&  18833&19.8& 623\phn($-$\phn 67,\phs$+$\phn 89)&\\
A1809............&   MP& 60&135025.2\phs052241&   1.66&  23696&37.1& 758\phn($-$\phn 65,\phs$+$\phn 86)&\\
                        &   MS& 54&135025.1\phs052241&   1.34&  23737& 4.7& 501\phn($-$\phn 35,\phs$+$\phn 40)&A1809\\
A1983............&   MP& 81&145038.2\phs165346&   1.82&  13471&$<0.1$& 634\phn($-$\phn 70,\phs$+$132)&\\
                        &   MS& 75&145038.2\phs165346&   1.70&  13492& 0.7& 514\phn($-$\phn 43,\phs$+$\phn 52)&A1983\\
                        &    S&  5&145058.4\phs165154&   0.23&  12606&50.7& 253\phn($-$\phn 42,\phs$+$130)&\\
A2052............&   MP& 51&151418.3\phs071335&   1.12&  10553&32.7& 641\phn($-$\phn 56,\phs$+$\phn 95)&\\
                        &   MS& 46&151416.8\phs071246&   1.09&  10459& 9.1& 561\phn($-$\phn 73,\phs$+$\phn 87)&A2052\\
A2063$-$..........&          MP&127&151911.5\phs075400&   6.76&  10934&$<0.1$&1404\phn($-$123,\phs$+$148)&\\
--MKW3S                      &  MS1& 91&152037.8\phs084742&   5.15&  10535& 4.4& 679\phn($-$\phn 46,\phs$+$\phn 49)&A2063\\
                        &    S&  7&151922.8\phs083508&   0.34&  11670&57.2& 997\phn($-$\phn 74,\phs$+$383)&\\
                        &  MS2& 26&151911.4\phs075356&   1.33&  13499&48.1& 603\phn($-$\phn 59,\phs$+$\phn 61)&MKW3S\\
A2107..............&MP=MS& 68&153729.6\phs215805&   1.00&  12337&64.0& 684\phn($-$\phn 60,\phs$+$\phn 79)&A2107\\
A2124..............&   MP& 62&154257.0\phs361433&   1.21&  19663&52.7& 872\phn($-$\phn 67,\phs$+$\phn 96)&\\
                        &   MS& 60&154259.0\phs361502&   1.21&  19619&20.9& 809\phn($-$\phn 60,\phs$+$\phn 73)&A2124\\
A2151..............&   MP&100&160311.4\phs175344&   1.61&  11034&51.3& 801\phn($-$\phn 46,\phs$+$\phn 64)&\\
                        &   MS& 98&160311.6\phs175344&   1.61&  11011&12.8& 762\phn($-$\phn 49,\phs$+$\phn 47)&A2151\\
                        &   S1& 19&160318.1\phs175413&   0.42&  10288&30.2& 644\phn($-$\phn 68,\phs$+$\phn 81)&\\
                        &   S2& 29&160337.3\phs181556&   0.66&  11259&79.0& 490\phn($-$\phn 47,\phs$+$\phn 74)&\\
                        &   S3&  5&160419.9\phs175429&   0.12&  11786&45.7& 219\phn($-$\phn 17,\phs$+$115)&\\
A2197$-$2199....&      MP& 78&162835.9\phs404423&  13.59&   9171&86.7& 686\phn($-$\phn 48,\phs$+$\phn 69)&\\
                        &   MS& 66&162835.9\phs404357&   3.59&   9094&20.8& 635\phn($-$\phn 41,\phs$+$\phn 65)&\\
                        &(MS1)& 37&162705.2\phs394240&   3.24&   9303&54.2& 686\phn($-$\phn 62,\phs$+$\phn 88)&A2199\\
                        & (S1)&  4&162918.5\phs395620&   0.17&   8883&94.0& 413\phn($-$100,\phs$+$149)&\\
                        &(MS2)& 30&162835.1\phs404502&   2.47&   8988&13.5& 585\phn($-$\phn 84,\phs$+$\phn 72)&A2197\\
                        & (S2)&  4&162654.2\phs411600&   0.10&   9924& 7.3& 695\phn($-$\phn 53,\phs$+$724)&\\
A2634$-$2666....&      MP&300&233606.5\phs264536&  13.76&   9074&$<0.1$&1409\phn($-$\phn 82,\phs$+$120)&\\
                        &   MS&264&233606.5\phs264536&   5.63&   9120& 0.9&1145\phn($-$\phn 63,\phs$+$\phn 82)&A2634=\\
                        &   S1& 26&234829.4\phs265735&   1.26&   8079&50.3& 386\phn($-$\phn 63,\phs$+$111)&=MS-S1-S2\\
                        &   S2& 22&233806.7\phs263409&   2.35&  11494& 4.9& 377\phn($-$\phn 25,\phs$+$\phn 58)&A2666\\
A2670..............&   MP&115&235139.9$-$104117 &   1.11&  22904& 5.4&1010\phn($-$\phn 70,\phs$+$\phn 85)&\\
                        &   MS&111&235139.9$-$104117 &   1.11&  22933&14.9& 918\phn($-$\phn 47,\phs$+$\phn 65)&A2670\\
A2670*............&    MP& 79&235140.0$-$104148 &   0.29&  22867&25.4&1103\phn($-$\phn 67,\phs$+$\phn 79)&\\
                        &   US&  8&235129.5$-$104210 &   0.10&  21096&57.9& 443\phn($-$\phn 61,\phs$+$153)&\\
                        &   MS& 68&235140.7$-$104151 &   0.30&  23026& 4.0& 912\phn($-$\phn 66,\phs$+$\phn 65)&\\
A2717..............&   MP& 53&240035.0$-$361229 &   1.26&  14715&47.7& 488\phn($-$\phn 38,\phs$+$\phn 49)&\\
                        &   MS& 52&240036.6$-$361219 &   1.26&  14703&11.0& 467\phn($-$\phn 35,\phs$+$\phn 38)&A2717\\
                        &    C& 18&240033.8$-$361257 &   0.47&  14276& 9.7& 364\phn($-$\phn 33,\phs$+$\phn 47)&\\
A2721..............&   MP& 83&000334.6$-$345928 &   1.61&  34356&$<0.1$&1092\phn($-$148,\phs$+$249)&\\
                        &   MS& 75&000335.0$-$345940 &   1.59&  34292&21.6& 841\phn($-$\phn 63,\phs$+$\phn 90)&A2721\\
\end{tabular}

\end{center}
\newpage
\begin{center}
\vspace{-3mm}
TABLE 3 {\it --- Continued}\\
\vspace{2mm}
\tiny
\begin{tabular}{llrrrrrrl}
\hline
\hline
\\
\multicolumn{2}{c}{Sample Name}
&\colhead{$N$}
&\colhead{Center 1950 ($\alpha$,$\delta$)}
&\colhead{$R$}
&\colhead{$<V>$}
&\colhead{$P_W$}
&\colhead{$\sigma$}
&\colhead{Identified}
\\
\multicolumn{2}{c}{}
&\colhead{}
&\colhead{}
&\colhead{(Mpc)}
&\colhead{(Km/s)}
&\colhead{}
&\colhead{(Km/s)}
&\colhead{Cluster}\\
\\
\hline
\\
A2877..............&   MP& 97&010734.9$-$461300 &   1.09&   7267& 2.7& 973\phn($-$\phn 77,\phs$+$\phn 82)&\\
                        &   MS& 86&010735.8$-$461308 &   1.04&   7111&51.3& 744\phn($-$\phn 51,\phs$+$\phn 63)&A2877\\
                        &    C& 16&010730.4$-$461430 &   0.14&   6779&62.5& 447\phn($-$\phn 42,\phs$+$\phn 65)&\\
                        &    S&  7&010645.0$-$460347 &   0.11&   7455&40.8& 345\phn($-$\phn 27,\phs$+$108)&\\
A3128..............&   MP&186&032927.9$-$524045 &   2.39&  17996&11.7& 869\phn($-$\phn 47,\phs$+$\phn 67)&\\
                        &   US& 22&032620.7$-$531801 &   1.54&  18006&27.3& 388\phn($-$\phn 74,\phs$+$\phn 95)&\\
                        &   MS&157&032927.5$-$524035 &   2.14&  17957&14.8& 841\phn($-$\phn 44,\phs$+$\phn 51)&A3128\\
                        &    C& 61&032929.6$-$524017 &   0.98&  17675& 1.9& 685\phn($-$\phn 42,\phs$+$\phn 54)&\\
A3266..............&   MP&130&043007.3$-$614030 &   1.33&  17811&78.3&1154\phn($-$\phn 67,\phs$+$\phn 92)&\\
                        &   US& 23&043215.8$-$612750 &   0.43&  17512&59.4& 528\phn($-$\phn 52,\phs$+$\phn 72)&\\
                        &   MS& 96&043026.8$-$613347 &   1.04&  17832& 8.3&1138\phn($-$\phn 74,\phs$+$\phn 94)&A3266\\
A3376..............&MP=MS& 77&060037.9$-$395622 &   2.29&  13909&95.3& 737\phn($-$\phn 57,\phs$+$\phn 88)&A3376\\
                        &    S& 11&060034.1$-$395642 &   0.92&  14156&18.1& 313\phn($-$\phn 39,\phs$+$117)&\\
A3391$-$3395....&      MP&211&062636.0$-$542435 &   2.64&  15424&$<0.1$&1241\phn($-$\phn 47,\phs$+$\phn 63)&\\
                        &  MS1&151&062632.2$-$542426 &   2.54&  14890& 2.4& 823\phn($-$\phn 43,\phs$+$\phn 51)&A3395\\
                        &   C1& 87&062627.9$-$542433 &   0.80&  15107&16.0& 740\phn($-$\phn 51,\phs$+$\phn 56)&\\
                        &  MS2& 53&062514.2$-$533951 &   2.64&  17081& 5.2& 786\phn($-$\phn 53,\phs$+$\phn 78)&A3391\\
                        &   C2& 29&062515.5$-$533941 &   0.75&  16476&18.1& 581\phn($-$\phn 40,\phs$+$\phn 73)&\\
A3526..............&   MP&112&124659.4$-$410644 &   1.49&   3623&$<0.1$& 930\phn($-$\phn 46,\phs$+$\phn 46)&A3526\\
                        &  MS1& 67&124636.1$-$410224 &   1.14&   3005& 0.4& 562\phn($-$\phn 34,\phs$+$\phn 54)&A3526A\\
                        &  MS2& 44&124936.3$-$410055 &   2.21&   4572&40.2& 294\phn($-$\phn 28,\phs$+$\phn 40)&A3526B\\
                        &    C& 15&124933.8$-$410105 &   0.46&   4749&60.8& 150\phn($-$\phn 19,\phs$+$\phn 37)&\\
A3526*............&    MP&102&124710.0$-$410844 &   1.68&   3533& 0.3& 883\phn($-$\phn 45,\phs$+$\phn 47)&\\
                        &  MS1& 69&124700.3$-$410442 &   1.48&   3063& 0.7& 520\phn($-$\phn 39,\phs$+$\phn 49)&\\
                        &    S&  8&124704.0$-$405238 &   0.58&   2078&73.1& 183\phn($-$\phn 24,\phs$+$\phn 50)&\\
                        &  MS2& 31&124928.2$-$410138 &   1.82&   4561&32.9& 249\phn($-$\phn 25,\phs$+$\phn 24)&\\
                        &    C&  6&124930.0$-$410255 &   0.45&   4856&57.8&  96\phn($-$\phn 10,\phs$+$\phn 28)&\\
Shapley............&   MP&482&132501.2$-$311324 &   4.61&  14277&$<0.1$&1075\phn($-$\phn 36,\phs$+$\phn 48)&\\
concentration            &  US1& 21&132056.8$-$312950 &   0.67&  14340&82.1& 583\phn($-$\phn 64,\phs$+$\phn 91)&\\
                        &  US2& 83&133208.6$-$312241 &   1.67&  14008& 0.3&1416\phn($-$134,\phs$+$125)&\\
                        &(US2S)& 20&133046.0$-$311739 &   0.57&  14098&42.7& 717\phn($-$\phn 70,\phs$+$\phn 87)&A3562\\
                        &   MS&373&132501.2$-$311324 &   2.59&  14292&12.4& 994\phn($-$\phn 33,\phs$+$\phn 45)&\\
                        &   S1& 44&132625.5$-$310215 &   0.71&  15044& 0.5& 755\phn($-$\phn 73,\phs$+$\phn 78)&\\
                        &   S2& 46&132822.5$-$313110 &   0.78&  13745&$<0.1$& 725\phn($-$\phn 50,\phs$+$\phn 85)&\\
                        &   S3& 95&132459.6$-$311307 &   0.95&  14320& 1.6& 735\phn($-$\phn 41,\phs$+$\phn 49)&A3558\\
A3667..............&   MP&163&200653.5$-$564955 &   2.66&  16664&59.7&1094\phn($-$\phn 55,\phs$+$\phn 81)&\\
                        &   MS&152&200653.5$-$564955 &   2.67&  16683&68.8&1052\phn($-$\phn 66,\phs$+$\phn 72)&A3667\\
                        &    S& 11&201002.4$-$570835 &   0.39&  15943&76.4& 487\phn($-$\phn 60,\phs$+$102)&\\
A3716..............&   MP& 92&204813.4$-$525820 &   2.20&  13684& 9.4& 843\phn($-$\phn 54,\phs$+$\phn 54)&\\
                        &   MS& 62&204813.3$-$525822 &   1.06&  13433& 9.6& 817\phn($-$\phn 47,\phs$+$\phn 65)&A3716S\\
A3888..............&   MP& 74&223130.2$-$375948 &   1.50&  45444& 1.0&1826\phn($-$180,\phs$+$248)&\\
                        &   MS& 64&223132.2$-$375938 &   1.53&  45682& 9.5&1307\phn($-$\phn 92,\phs$+$100)&A3888\\
A4038..............&MP=MS& 43&234505.4$-$282443 &   0.46&   8630& 1.0& 898\phn($-$116,\phs$+$112)&A4038\\
\\
\hline
\end{tabular}

\end{center}
{\footnotesize\parindent=-3.5mm
NOTE. -- The bracketed structures are detected at a
c.l. $<99.5\%$. The samples marked by an asterisk are alternative initial
samples, with a lower completness level or a smaller extension.}

\vspace{-3mm}
\normalsize
\begin{multicols}{2}

This suggests that we are considering a
region large enough so that the effects of (possible) velocity
anisotropies are already averaged and hence the global value of galaxy
velocity dispersion is independent of possible velocity anisotropies.
Therefore, we take the VDP-value corresponding to the external region
as our estimate of $\sigma$.  In this paper we interpret the observed
behaviours of the VDP in internal region as being due to the presence
of velocity anisotropy although they could be explained also by
peculiarities of internal relative distribution of mass and galaxies
(e.g. Merritt 1987).

\section{General Results and Discussions}

Our fairly large sample of clusters enables us to draw some
general results in a statistical way.  In the following
analyses, we do not consider clusters A2197, A2199, and A3562,
which are identified with a small statistical significance, and 
cluster A539, whose internal structure is not clearly understood.

These analyses concern the ``identified clusters'' which, in 21
cases, do not correspond to the main peak.
For these 21 clusters, our cluster
identification always leads to a reduction in the value of $\sigma$;
the $\sigma$ of the main peak shows a mean overestimate of about
$18\%$, with a maximum of $58\%$ for cluster A1185.

In general, after the cluster identification, the velocity dispersion
profile (VDP) becomes less noisy and flatter in the outer regions.
The VDP drastically changes in the case of fields which contain
more than one system (see Figure 2).

\subsection{Morphological classification}

One of our main aims is the classification of the structure of galaxy
clusters in order to better understand their morphology.  In Table~4
we attempt a classification based on the substructures detected within
each cluster area.  We only classify the samples identified as {\em
identified clusters}, without considering the initial cluster fields whose
morphologies depend on the extension of the observed area.

We introduce some morphological categories, which depend on the
cluster appearance at different scales.

We define three categories:

{\em  Unimodal}: these clusters appear as single systems at large
scales.

{\em Bimodal}: these clusters (A754, A548, A3526) show two main
 systems at large scales. Moreover, we consider also A1736 and A3716
 to be bimodal clusters as they are well known to be bimodal in the
 literature, although in this paper we analyzed only one of their
 components.

{\em Complex} : these clusters show substructures which involve a
large part of the main system itself (cluster A2151 and the clump
denominated A548SW).

For clusters analyzed, we list in Table~4 the presence of
substructures and/or any kind of irregularity detected in our analysis
( i.e. a non-Gaussian velocity distribution, $P_W < 5\%$, and/or a cD
galaxy with peculiar velocity). \\

\end{multicols}
\includegraphics{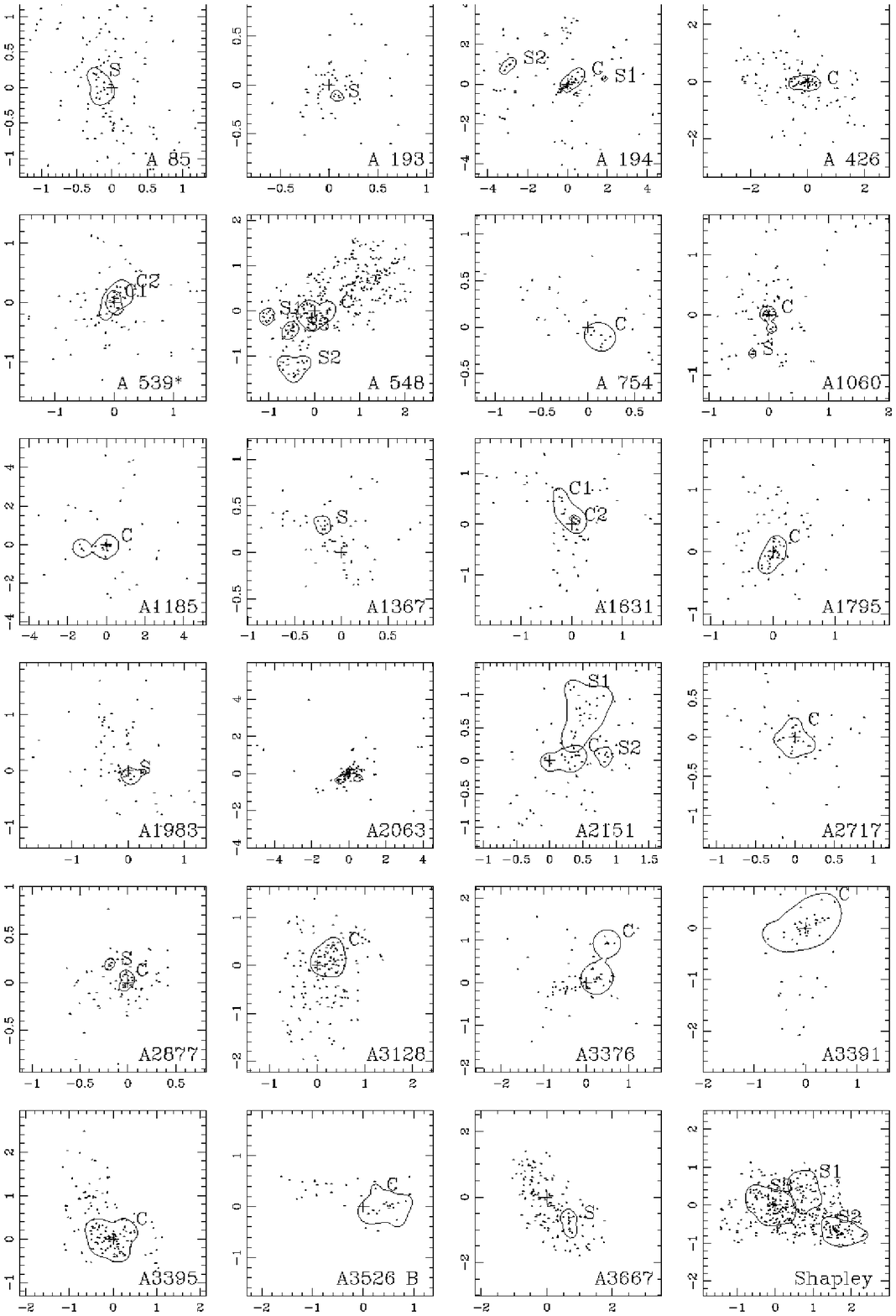}
$\ \ \ \ \ \ $\\
\vspace{22.5cm}
$\ \ \ $\\
\vspace{1mm}
{\small\parindent=3.5mm
{\sc Fig.}~3.---
Spatial distribution of galaxies of {\em  identified
clusters}, which have internal structures, are shown.
Contours contain all the members assigned to respective substructures.
X-ray centers  are indicated by crosses.
}
\vspace{5mm}

\begin{center}
\begin{minipage}{14cm}
\renewcommand{\arraystretch}{1.0}
\renewcommand{\tabcolsep}{3.5mm}
\begin{center}
\vspace{-3mm}
TABLE 4\\
\vspace{2mm}
{\sc Morphological Classification\\}
\vspace{2mm}
\scriptsize
\begin{tabular}{llc|llc}
\hline
\hline
\multicolumn{3}{c|}{}\\
\colhead{Cluster}&
\colhead{Morphology}&
\multicolumn{1}{c|}{Irregularities}&
\colhead{Cluster}&
\colhead{Morphology}&
\colhead{Irregularities}\\
\colhead{(1)}&
\colhead{(2)}&
\multicolumn{1}{c|}{(3)}&
\colhead{(1)}&
\colhead{(2)}&
\colhead{(3)}\\
\multicolumn{3}{c|}{}\\
\hline
\multicolumn{3}{c|}{}\\
A0085         & Uni & sub, $P_W<5$\%          &A2052         & Uni & \nodata                 \\
A0119         & Uni & \nodata                 &A2063         & Uni & sub, pec. cD, $P_W<5$\%\\
A0151         & Uni & \nodata                 &A2107         & Uni & pec. cD                 \\
A0193         & Uni & sub                     &A2124         & Uni & \nodata                 \\
A0194         & Uni & sub                     &A2151         & Comp& sub                     \\
A0262         & Uni & \nodata                 &A2634         & Uni & $P_W<5$\%               \\
A0399         & Uni & $P_W<5$\%               &A2666         & Uni & \nodata                 \\
A0401         & Uni & pec. cD                 &A2670         & Uni &  pec. cD                \\
A0426         & Uni & \nodata                 &A2717         & Uni & \nodata                 \\
A0548         & Bi  & \nodata                 &A2721         & Uni & \nodata                 \\
\phm{A0548}SW & Comp& sub, $P_W<5$\%          &A2877         & Uni & sub                     \\
\phm{A0548}NE & Uni & $P_W<5$\%               &A3128         & Uni & \nodata                 \\
A0754         & Bi  & \nodata                 &A3266         & Uni & \nodata                 \\
\phm{A0754}NW & Uni & $P_W<5$\%               &A3376         & Uni &  \nodata                \\
\phm{A0754}SE & Uni & $P_W<5$\%               &A3391         & Uni & pec. cD                 \\
A1060         & Uni & sub                     &A3395         & Uni & \nodata                 \\
A1146         & Uni & pec. cD                 &A3526         & Bi  & \nodata                 \\
A1185         & Uni & $P_W<5$\%               &\phm{A3526}A  & Uni & $P_W<5$\%               \\
A1367         & Uni & sub                     &\phm{A3526}B  & Uni & \nodata                 \\
A1631         & Uni & \nodata                 &A3558         & Uni & pec. cD, $P_W<5$\%      \\
A1644         & Uni & pec. cD                 &A3667         & Uni & sub                     \\
A1736         & Bi  & \nodata                 &A3716         & Bi  & \nodata                 \\
\phm{A1736}A  &\nodata& \nodata               &\phm{A3716}N  &\nodata& \nodata               \\
\phm{A1736}B  & Uni & $P_W<5$\%               &\phm{A3716}S  & Uni & \nodata                 \\
A1795         & Uni & \nodata                 &A3888         & Uni & \nodata                 \\
A1809         & Uni & $P_W<5$\%               &A4038         & Uni & $P_W<5$\%               \\
A1983         & Uni & sub                     &MKW3S         & Uni & \nodata                 \\
\multicolumn{3}{c|}{}\\
\hline
\end{tabular}

\end{center}
{\footnotesize\parindent=-3.5mm
 NOTE. --- ``Uni'', ``Bi'', ``Comp'', mean unimodal, bimodal, and
complex, respectively. ``Sub'' indicates the
presence of substuctures, ``pec. cD'' the presence of a cD galaxy with
peculiar velocity, and ``$P_W<5$\%'' a low probability of Gaussianity of
the velocity distribution.}

\vspace{0mm}
\end{minipage}
\end{center}
\normalsize
\begin{multicols}{2}

Here a cD galaxy is defined as having
a {\em peculiar} velocity by adopting the robust test by Gebhardt \&
Beers (1991), and considering the 95\% c.l.

For a limited number of clusters we made a comparison with the results
obtained by Gurzadyan \& Mazure (1996) by means of a recently
developed method, which enables one to study the hierarchical
properties of the subsystems by taking into account the positions,
redshifts and magnitudes of cluster galaxies, and thus to assign the
full system membership (see Gurzadyan, Harutyunyan, \& Kocharyan
1994).  For six clusters in common with our sample, there appears to
be fair agreement in the identification of the main system and of the
most prominent substructures.

Out of 44 clusters, we classify five clusters as bimodal, one as
complex, and the others as unimodal.  In 9 of the 38 unimodal clusters
we clearly detect the presence of small-scale substructures and there
is some sign of them in 12 others. Hence, we detect substructures in about
one third (15/44) of our clusters. This is in broad agreement with previous
statistical works which employ different techniques (Geller \& Beers 1982;
Dressler \& Schectman 1988b; Jones \& Forman 1992).

Our cluster sample is, however, slightly biased towards more regular
clusters. In fact about half of our clusters are cD ones, which are
usually better studied, while we verified that in Northen Abell
Catalog only $\sim$ 20\% of the nearby (Abell distance class $\leq 4$)
clusters are classified as cD.  Hence, the result of our
classification may not strictly be representative of the Universe.

\subsection{Optical and X-ray results}

In the detailed discussion of individual clusters (see the appendix) and in
Figure~4 we have often compared our results obtained from optical data
with those coming from X-ray studies.  Here we may summarize some main
points.  We consider 42 clusters or clumps for which there is a
corresponding unambiguous identification in X-ray maps. Therefore we
do not consider clusters A754, A2151, A3526.  We found that the mean
distance between X-ray and optical centers is 0.11 \h, which is
roughly the typical uncertainty in the estimate of cluster centers
(e.g. Beers \& Tonry 1986; Rhee \& Latour 1991).

In our sample the mean (absolute) percent difference between the
$\sigma$-value and the corresponding $T$-value is about $17\%$.  This
discrepancy is consistent with typical errors on $\sigma$ ($8\%$), and
on $T$ ($12\%$), for 29 clusters having a direct measure of $T$. The
other clusters, whose $T$ is estimated by X-ray luminosity, are
supposed to be affected by larger errors on $T$.  Figure~4 shows that
$\sigma$ and $T$  agree well for most clusters, with two exceptions
(A119 and A1060), whose measures differ by more than two s.d..  For
A2634 the agreement is acceptable within the presumably virialized
region.  The general good agreement between global X-ray and optical
cluster properties make us confident in assuming dynamical equilibrium
of both galaxy and gas components within the cluster potential and
thus in using the virial mass estimator. The mean value of $\beta$ is
$0.90$ with a rms = 0.29.

\end{multicols}
\includegraphics{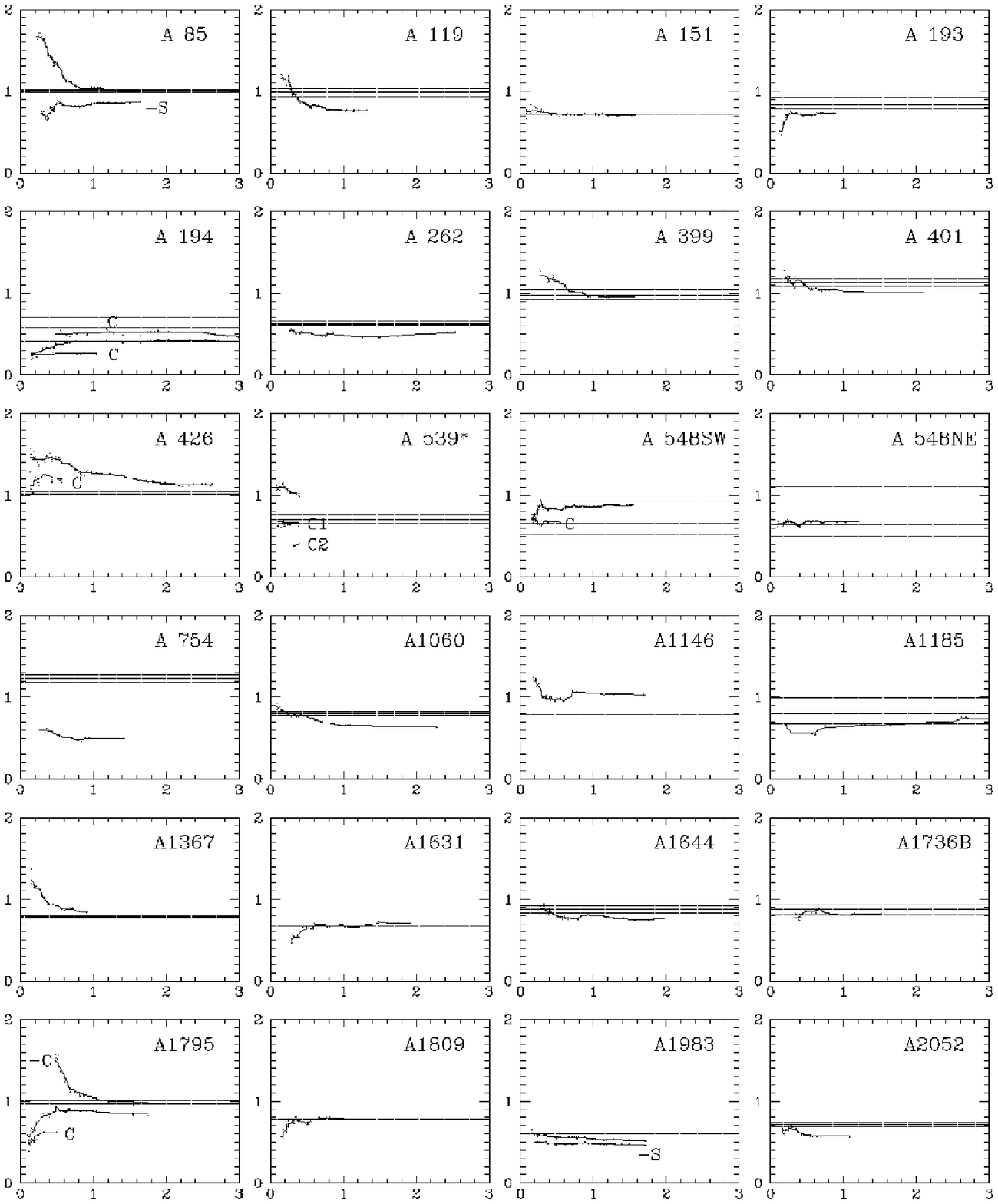}
$\ \ \ \ \ \ $\\
\vspace{20.5cm}
$\ \ \ $\\
\vspace{1mm}
{\small\parindent=3.5mm
{\sc Fig.}~4.---
The velocity dispersion profiles (VDP), where
the dispersion at a given radius is the average l.o.s. velocity
dispersion within this radius. The VDP units are $10^3$ \ks, and
distance from the center is expressed in Mpc. The center is generally
the X-ray center if not specified otherwise in the text. The first
point represents the $\sigma$ as computed for the ten galaxies, which
are closest to the cluster center. Subsequently points are computed
considering one more galaxy each time.  A smoothed line is also
superimposed.  The horizontal lines show the values of $\sigma$, and
the respective error bands obtained from the X-ray temperatures (see
Table~1) on the condition of perfect galaxy/gas energy equipartition,
i.e. $\beta = 1$.
}
\vspace{5mm}
\newpage
\includegraphics{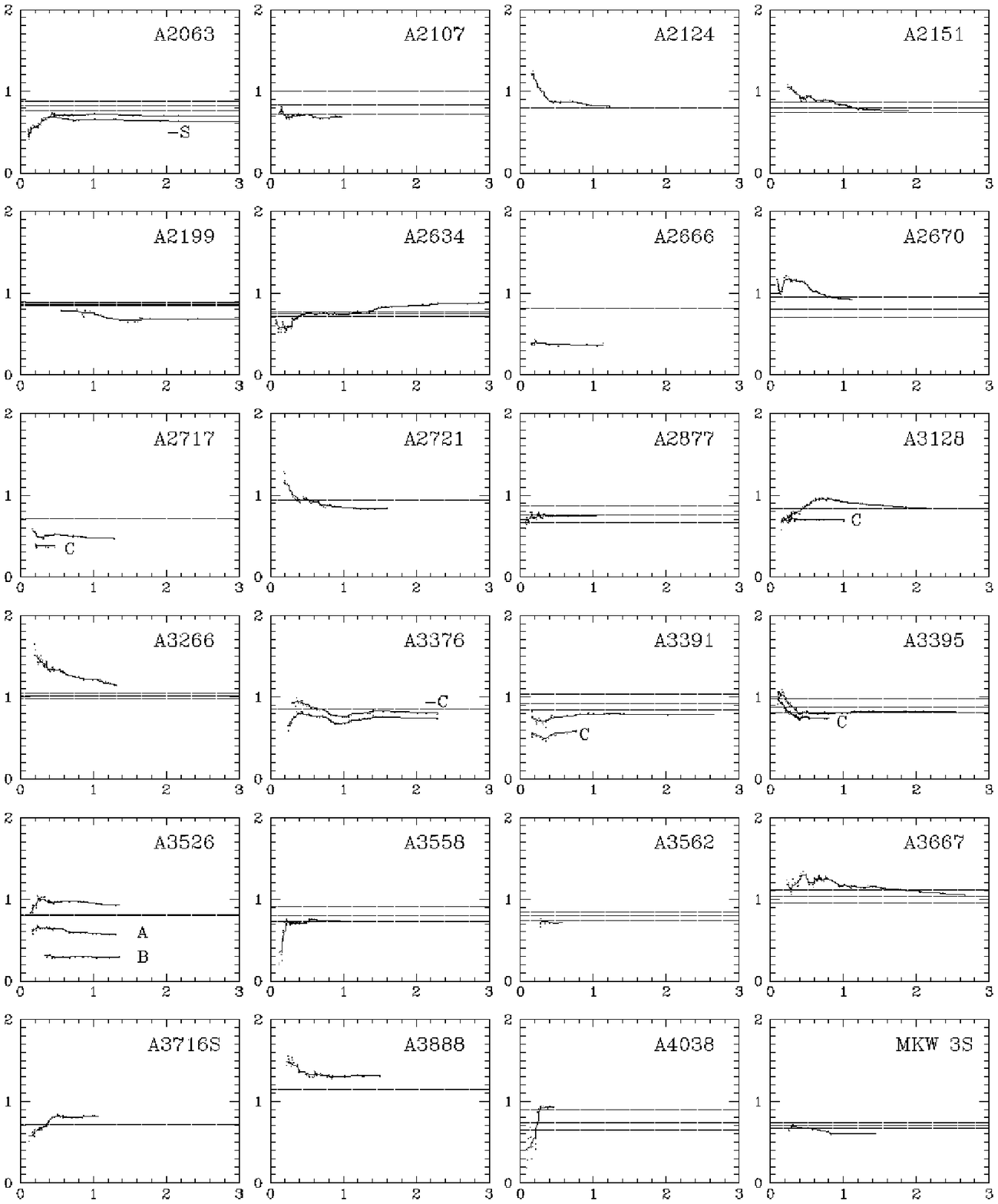}
$\ \ \ \ \ \ $\\
\vspace{20.5cm}
$\ \ \ $\\
\vspace{-11mm}
\begin{center}
{\sc Fig.}~4.--- Continued.
\end{center}
\vspace{-5mm}
\begin{multicols}{2}

\subsection{Unimodal Clusters}

In the case of 9 unimodal clusters, which show the presence of
substructures, we analyzed the effect of substructures on the
kinematics and dynamics of clusters by comparing the values of
$\sigma$ and mass computed before and after rejection of the detected
substructures. We adopted the {\em standard } virial mass (see
e.g. Giuricin, Mardirossian, \& Mezzetti 1982), which is strictly
valid only if the mass distribution follows galaxy distribution
(e.g. Merritt 1987; Merritt 1988).  However, the same hypothesis is
also assumed in other usual mass estimators, e.g. in the projected
mass estimator (Heisler, Tremaine \& Bahcall 1985) used by B95, whose
results will be compared with ours.  
\end{multicols}
\renewcommand{\arraystretch}{1.0}
\renewcommand{\tabcolsep}{3.5mm}
\begin{center}
\vspace{-3mm}
TABLE 5\\
\vspace{2mm}
{\sc The Effect of Substructures\\}
\vspace{2mm}
\scriptsize
\begin{tabular}{lrrrrcrr}
\hline
\hline
\\
\colhead{Cluster}
&\multicolumn{2}{c}{$\sigma$ (Km/s)}
&\multicolumn{2}{c}{Mass$_{<1.5 Mpc}$ ($10^{14} M_{\odot}$)}
&\colhead{$R_v$ (Mpc)}
&\multicolumn{2}{c}{Mass$_{<R_v}$ ($10^{14} M_{\odot}$)}
\\
\colhead{}
&\colhead{before}
&\colhead{after}
&\colhead{before}
&\colhead{after}
&\colhead{}
&\colhead{before}
&\colhead{after}
\\
\colhead{(1)}
&\colhead{(2)}
&\colhead{(3)}
&\colhead{(4)}
&\colhead{(5)}
&\colhead{(6)}
&\colhead{(7)}
&\colhead{(8)}\\
\\
\hline
\\
A0085 & 995$(-77, +88)$& 853$(-47, +59)$& 13.4$\pm$2.1& 6.5$\pm$1.1\phm{00}& 1.66&13.4$\pm$2.1&10.8$\pm$1.5\\
A0193 & 723$(-58, +90)$& 751$(-64, +78)$&  3.6$\pm$0.9& 4.1$\pm$0.9\phm{00}& 1.12& 3.6$\pm$0.9& 4.1$\pm$0.9\\
A0194 & 389$(-45, +54)$& 389$(-45, +54)$&  1.4$\pm$0.4& 1.4$\pm$0.4\phm{00}& 0.54& 0.8$\pm$0.3& 0.8$\pm$0.3\\
A1060 & 639$(-39, +49)$& 649$(-42, +49)$&  3.8$\pm$0.5& 3.8$\pm$0.6\phm{00}& 1.04& 3.5$\pm$0.5& 3.5$\pm$0.5\\
A1367 & 836$(-69, +79)$& 881$(-59, +80)$&  5.2$\pm$1.1& 6.2$\pm$1.0\phm{00}& 0.99& 5.2$\pm$1.1& 6.2$\pm$1.0\\
A1983 & 528$(-25, +60)$& 472$(-31, +56)$&  3.1$\pm$0.6& 2.6$\pm$0.6\phm{00}& 0.59& 1.8$\pm$0.4& 1.2$\pm$0.3\\
A2063 & 701$(-51, +68)$& 639$(-47, +72)$&  4.0$\pm$0.8& 3.1$\pm$0.7\phm{00}& 1.10& 3.7$\pm$0.8& 2.9$\pm$0.7\\
A2877 & 741$(-55, +59)$& 766$(-59, +62)$&  3.2$\pm$0.6& 3.5$\pm$0.7\phm{00}& 0.94& 3.1$\pm$0.6& 3.3$\pm$0.7\\
A3667 &1141$(-76, +39)$&1092$(-70, +86)$& 17.1$\pm$2.4&17.0$\pm$2.5\phm{00}& 1.74&17.5$\pm$2.4&18.4$\pm$2.6\\ 
\\
\hline
\end{tabular}

\end{center}
\normalsize
\begin{multicols}{2}

The virial mass estimate does not
require any assumptions about the isotropy of galaxy orbits, since
each possible effect vanishes by averaging the velocity dispersion
over the whole cluster sample (Merritt 1987).  The virial mass is
fully meaningful when it is computed within the virialization radius,
which corresponds to the region where the cluster is expected to have
reached dynamical equilibrium at the present epoch.  The virialization
radii, $R_v$, were computed in the same way as B95, i.e. by assuming a
proportionality between the X-ray temperature and $R_v^2$ and by
scaling to Coma values.

In Table~5 we list several quantities computed before and after the
rejection of galaxies, which belong to the detected substructures: in
Cols.~(2) and (3) the global $\sigma$ (in \ks); in Cols.~(4) and (5) the
virial mass as computed within the Abell radius (i.e. 1.5 \h); in
Col.~(6) the virialization radius in \h; in Cols.~(7) and (8) the virial
mass as computed within the virialization radius.

The $\sigma$ and mass distributions computed before and after removing
substructures are not different according to the Kolmogorov-Smirnov
test (see Press et  al. 1992).  This holds true both for masses and for
$\sigma$ computed within a fixed radius of 1.5 \h and within the
virialization radius. As far as the individual values are concerned,
only the mass of A85 shows a significant change.  Indeed, as discussed
in the appendix, its substructure is likely to be unbound and thus it may be
only a chance superimposition on the cluster.

 Our results are in agreement with E94 and partially in disagreement
with B95, who found that the masses computed within the fixed radius
depend on a possible correction for the presence of substructures.
This is probably due to the fact that we consider as a {\em identified
cluster}, in which we look for substructures, a sample detected by a
more refined method than that used by B95, who employed a simple cut
in velocity space within an Abell radius (Bird 1994). This fact could
also explain why she detected substructures in clusters which we found
were not substructured: the galaxies which belong to these
substructures have probably been excluded in our procedure of
identification of the cluster. Indeed, also B95 found no
significant effect when the masses are computed within the
virialization radius; in fact, this radius is generally smaller than
1.5 \h and therefore the galaxy sample is naturally better
cleaned. 

  Reassuringly, by considering the 16 unimodal clusters in
common with B95, our distribution function of mass values does not
significantly differ from that of B95's mass values corrected for
substructures (column 10 in Table 1 of B95), according to the
Kolmogorov-Smirnov test.

Our results do not disagree with Pinkney et al. (1996), who found, by
considering simulations of a cluster with a merging clump (mass ratio
1:6 and 1:3), that the virial mass estimator overestimate the true
cluster mass, in particular in the case of small projection angles of
merger axis.  In fact, our substructures are generally smaller (see
Table~6) and the projection angles are likely greater. The most
probable case of head-on merging, A85, actually shows a significant
change in mass estimation.

Substructures can significantly affect small scale phenomena,
e.g. the peculiarity of the cD galaxy velocities which
depends on the removal of some substructures (see, e.g., A2063).

In Table~6 we list some interesting parameters for each substructure
contained in unimodal clusters: in Col.~(2), $N_{\%}$, the fraction of
galaxies relative to the total of the galaxies contained within 1 \h;
in Col.~(3), $R$, the maximum projected radius of the substructure
computed by considering the biweight center (which is preferable to
the density center when the number of galaxies is small, see Beers et
al. 1990).  The average values are $N_{\%}=7.5\%$ and $R=0.21$ \h.

The core structures detected do not appear to be a homogeneous class:
they can contain few or many galaxies and can have a lower or higher
$\sigma$ than that of the whole cluster. The statistics is still too
poor to  draw general conclusions.  However, they are so
close in position and velocity to the respective cluster quantities,
that we suspect they are cluster regions with a particular
kinematical status (see e.g. A1795) or, if very extended, real
virialized clusters (see e.g. A1185, A3391), rather than, e.g., the
relics of some structures coming from the outside.  Therefore, in the present
analysis we do not consider samples obtained by rejecting the galaxies
of core structures because it is probable that the remaining
structures have no real physical significance.

\hspace{-4mm}
\begin{minipage}{9cm}
\renewcommand{\arraystretch}{1.2}
\renewcommand{\tabcolsep}{3.mm}
\begin{center}
\vspace{-3mm}
TABLE 6\\
\vspace{2mm}
{\sc The Substructures\\}
\vspace{2mm}
\footnotesize
\scriptsize
\begin{tabular}{lll}
\hline
\hline
\colhead{Cluster}
&\colhead{$N_{\%}$}
&\colhead{$R$}\\
\hline
A0085S   &  8&    0.19 \\
A0193S & 7 &    0.03 \\  
A0194S1 &7  &    0.11 \\
A0194S2 &17 &    0.34 \\
A1060S& 6 &    0.04 \\
A1367S& 9 &    0.08 \\
A1983S& 9 &    0.22 \\
A2063S& 10&    0.27 \\
A2877S& 8 &    0.10 \\
A3667S& 13&    0.30 \\
\hline
\end{tabular}

\end{center}
\vspace{0mm}
\end{minipage}
\normalsize

Nine of our clusters (A193, A194, A1631, A1795, A1809, A2063, A3128,
A3558, A4038) show velocity dispersion profiles which decrease towards
the cluster center.  This behaviour may be due to the effect of
dynamical friction, which slows down the most luminous, central
galaxies with respect to the background matter (Merritt 1988), or to
the loss of orbital energy during galaxy merging (Menci \&
Fusco-Femiano 1996).  The same processes could explain the presence of
core structures with a low-$\sigma$ population which we find in four
of the above clusters (A194, A1631, A1795, A3128) and in another
cluster (A2877).  The final consequence of these processes could be
the formation of a cD galaxy. Nevertheless, only five of the
above-mentioned clusters are cD clusters (A193, A1795, A1809, A2063,
A3558) as expected in our sample, where about half of the clusters are
cD clusters.

The above-mentioned behaviour of VDP as well as the core structures
with a low velocity dispersion could be also explained by alternative
scenarios. The velocity dispersion could
increase with radius because of the inclusion of subclumps with different
mean velocities or dispersions. However, some of the above
clusters show no substructures in outer regions (e.g. A1631, A1795). In
another scenario, the low dispersion population could be the
remnants of a small subcluster, as suggested for the A576 cluster by
Mohr et al. (1996).

The presence of a cooling flow and/or luminosity segregation,
which are signs of possible relaxation, could allow us to distinguish
whether the observed effects are due to dynamical relaxation or to the
presence of substructures but, at present, the available information is
poor.  For instance, A1795 is well known to have a strong cooling flow,
A2063 has a faint one, (Edge, Stewart, \& Fabian 1992), as does
A3558 (Bardelli et al. 1996), and A1809 has no cooling flow (Stewart
et al. 1984).  Biviano et al (1992) found evidence that luminous
galaxies are segregated in velocity in A194, but they did not find any
significant velocity segregation for A1631, A3128, A3558.
Den Hartog \& Katgert (1996)  found faint evidence of luminosity 
segregation in A3128 and A3558 and no evidence in A194, A1631, A1809,
and A2063.

\subsection{Bimodal and Complex Clusters}

As regards  bimodal or complex clusters, the VDPs (see Figure~4 and
Table~3) suggest that their internal kinematics strongly biases the
estimate of $\sigma$. 

\hspace{-4mm}
\begin{minipage}{9cm}
\renewcommand{\arraystretch}{1.2}
\renewcommand{\tabcolsep}{3.mm}
\begin{center}
\vspace{-3mm}
TABLE 7\\
\vspace{2mm}
{\sc Bimodal and Complex Clusters\\}
\vspace{2mm}
\footnotesize
\scriptsize
\begin{tabular}{llr}
\hline
\hline
\colhead{Cluster}
&\colhead{Sample}
&\colhead{Mass ($10^{14} M_{\odot}$)}
\\
\colhead{(1)}
&\colhead{(2)}
&\colhead{(3)}\\
\hline
A0548 .................& MP & 9.5$\pm$0.9\phm{0000}\\ 
\phm{A0548}&MS1+MS2&13.0$\pm$1.3\phm{0000}\\
\phm{A0548}&C+S1+S2+S3+MS2&9.5$\pm$2.0\phm{0000}\\
\phm{A0548}&C+MS2&6.4$\pm$0.9\phm{0000}\\
A0754 .................& MS & 1.7$\pm$0.6\phm{0000}\\ 
\phm{A0754}&MS1+MS2&2.5$\pm$0.9\phm{0000}\\
A2151 .................&MS & 6.0$\pm$0.9\phm{0000}\\
\phm{A2151}&S1+S2+S3&4.3$\pm$1.7\phm{0000}\\
A3526 .................& MP & 7.5$\pm$0.9\phm{0000}\\ 
\phm{A3526}&MS1+MS2&2.3$\pm$0.5\phm{0000}\\
\hline
\end{tabular}

\end{center}
\vspace{3mm}
\end{minipage}
\normalsize

In fact, these clusters could be cases of
ongoing merging and their dynamical status may be rather far from
virial equilibrium. In this respect, we stress the importance of using
optical information, when one suspects a strong cluster merging and
the presence of compression-heated gas (e.g. Zabludoff \& Zaritsky
1995).  In fact, the optical component seems much less disturbed by
cluster collisions than the gas content, so that the galaxy systems
could survive the first cluster encounter (McGlynn \& Fabian
1984). In these cases the most meaningful mass estimate could thus be
the sum of  the virial masses of the  (supposed virialized) clumps.

Table~7 shows, in Col.~(3), the cluster masses computed both as the virial
mass of the whole cluster and as the sum of virial masses of each
subclump.  Indeed, a precise mass estimate depends on the choice of
the clumps considered (see, e.g, several mass estimates of A548).  All
these masses are computed within the respective virialization radii
(\S~5.3), which are here computed by adopting $\sigma$ rather than the
X-ray temperature as an estimate of the potential depth. This choice
is due to the fact that some systems are not clearly spatially
identified in X-ray maps and that observed X-ray temperatures could
not reliably measure the cluster potential well. 

These two ways of computing the mass may give appreciably different
results.  In particular, in the case of the {\em head-on } bimodal
cluster A3526, neglecting the presence of clumps in velocity space
leads one to strongly overestimate the cluster mass. Indeed, by
analyzing the two-clumps merging in simulated clusters,
Pinkney et al. (1996) found that the cluster mass is strongly
overestimated in the case of {\em head-on } two-clumps merging.

More accurate mass evaluations need the development of hydrodynamical
simulations, which include a large range of initial parameters
(e.g. angles of view and encounter velocities) and which describe both
collisional and acollisional cluster components (e.g. Burns et
al. 1995).

\section{Summary and Conclusions}

We analyzed a set of 48 galaxy clusters, which is the most extensive
sample in the literature used to study the presence of substructures
by means of galaxy positions and redshifts.  

We used a multi-scale analysis which couples kinematical estimators
with the wavelet transform (Escalera \& Mazure 1992; Escalera et
al. 1994), by introducing three new kinematical estimators.  These
estimators parameterize the departures of the local means and/or local
dispersions of the measured radial velocities with respect to their
global values for the environment. 

Both the methods we apply for detecting substructures and
for computing velocity means and dispersions (Beers et al. 1990)
have the advantage of requiring no Gaussian velocity distributions.
In fact, one expects non-Gaussian galaxy velocity distributions
in clusters that, even in dynamical equilibrium, show the presence of
velocity anisotropy in galaxy orbits (e.g. Merritt 1987).

We analyzed 44 cluster fields, recovering 48 clusters, of which 44 are
detected with high significance (99.5\%) and are sampled up to a
sufficiently external region.  Of the 44 clusters, we classify five
clusters as bimodal, one as complex, and the others as unimodal.  In 9
of the 38 unimodal clusters we clearly detect the presence of
small-scale substructures and there is some sign of them in 12 others.
Hence, we detect the presence of substructures in about one third of
the clusters, in broad agreement with previous works which are based on
different techniques. However, the high fraction
of cD clusters in our sample (about 50\%) suggests that our sample may
be not strictly representative of the Universe.  Indeed, this is the
first part of a larger study planned to consider the other clusters
specific to the ENACS database (Katgert et al. 1996) in order to
obtain a more statistically significant sample.

To discuss the effect of substructures on cluster dynamics, one should
consider that substructures can assume some basic forms (West \&
Bothun 1990).

The groups, which are not dynamically bound to the cluster, or bound
units, which reside outside the relaxed portion of the cluster and are
perhaps just falling in, are probably rejected in our phase of galaxy
cluster identification, with the possible exception of A85.

The dynamical substructures that reside within an otherwise relaxed
system may be the remnants of a previous secondary infall or cluster
merging.  The small-scale substructures we detected represent, on
average, 7.5\% of the cluster galaxies within 1 \h and their average
extension is $\sim$ 0.2 \h. The two values are in agreement with
typical population fractions and sizes of substructures inferred by
small-scale correlations among galaxies observed in many apparently
relaxed clusters (Salvador-Sol\'e, Gonz\'alez-Casado, \& Solanes 1993;
Gonz\'alez-Casado, Solanes, \& Salvador-Sol\'e, 1993).  The
substructures we detect are probably sufficiently compact to survive
the cluster force after merging, according to the theoretical work by
Gonz\`ales-Casado et al. (1994).  These authors have suggested that
these substructures could be the remnants of massive cores of groups
or small clusters.

The effect of small substructures does not appear to be considerable on
the global cluster kinematics and dynamics, i.e.  on the value of the
velocity dispersion and mass. This indicates that clusters which
show only small substructures are not too far from dynamical
equilibrium, as is also suggested by the generally good agreement between
global X-ray and optical cluster properties (centers and velocity
dispersions).

The above conclusions do not hold true for bimodal or complex clusters,
which are likely cases of recent cluster merging.

  From the point of view of statistical studies concerning galaxy
clusters, the problem of the estimate of velocity dispersion and mass
in bimodal and complex clusters might not be serious if their fraction
is fairly small as in our sample.  This could explain the result
obtained by Biviano et al. (1993), who found no difference between the
mass distribution of substructured and non-substructured clusters, and
by Fadda et al. (1996), who found no difference in the cumulative
distributions of cluster velocity dispersions whether or not they took
into account the multimodality of some clusters in their velocity
distributions.

\acknowledgements 

We wish to thank the ENACS team for providing us with new data prior
to publication.  We thank the anonymous referee for useful remarks and
comments.  We are also indebted to Harald Ebeling, who gave us some
X-ray data in advance of publication.  We are grateful also to Vahe
Gurzadyan for some enriching discussions on the philosophy of
structure detection.

This work has been partially supported by the {\em Italian Ministry of
University, Scientific Technological Research (MURST)}, by the {\em
Italian Space Agency (ASI)}, and by the {\em Italian Research Council
(CNR-GNA)}.

\end{multicols}
\appendix
\section{Results for Individual Clusters.}

We organize the presentation in the form of a series of paragraphs,
each one corresponding to a cluster.  For each cluster we describe the
pure detection results obtained from our main procedure of systems
identification.  Moreover, by performing some particular analyses, as
well by comparing our findings with the relevant results in the
literature, we suggest the most probable dynamical status. It should be
noted that it is not the purpose of this work to provide definitive
conclusions regarding the dynamics of these clusters.

In the following discussions some notations are used.

The word {\em  regular} means an almost
symmetrical spatial shape combined with a Gaussian or nearly Gaussian velocity
distribution.

To test whether two systems are unbound, we apply the two-body model
(e.g. Beers, Geller \& Huchra 1982), which gives both bound and unbound
solutions by varying the value of the unknown projection angle between
the two systems. When no bound solutions are possible (Newton's
criterion), we classify two structures as {\em  unbound}.
\\

{\em  A85. ---} Regular shaped cluster which contains a foreground
group of 7 galaxies (S) in front of the center.  No further
substructure.  The foreground group roughly corresponds to the one
already shown by Malumuth et al. (1992) and Beers et al. (1991).
Newton's criterion does not exclude the possibility that the S group
may be bound. However, even if it is bound, it could be a sign of
secondary infall to a pre-virialized cluster.  Removing this group
changes the $\sigma$ (877$^{+61}_{-50}$ \ks), which becomes lower than
the $\sigma$ expected from $T$, although still consistent at about two
s.d..  Moreover, the cleaned cluster shows an acceptably Gaussian
velocity distribution.

{\em  A119. ---} Regular shaped cluster.  
From inspection of X-ray maps and galaxy isopleths derived from
photometry, Fabricant et al. (1993) suggested the presence of
multiple structures. Although the number of galaxy redshifts
is now almost doubled, the cluster kinematics does not show any
evidence of substructures.  This is not, however, a clear contradiction
of the suggestion of Fabricant et al. (1993), since their supposed
configuration is beyond the limits of our detection method (see \S3.1).  
The only possible evidence of substructures is the disagreement
between  $\sigma$ and $T$ (at about 2.5 s.d.).

{\em  A151. ---} Two distinct populations in terms of velocity
(MS1,MS2), separated by almost 4000 Km/s, slightly overlapping but
easy to identify.  These two populations correspond to those already
identified by Proust et al. (1992): the real cluster and a
foreground group, respectively.  We confirm that these systems are
unbound according to Newton's criterion.

{\em  A193. ---} Regular cluster.  However, a close quartet close to the
center appears significant (S).  The VDP shows a strong decrease
towards the cluster center, which suggests a possible advanced
dynamical status (see the discussion in \S~5.3), confirmed also by the
presence of a cD galaxy.

{\em  A194. ---}  Two loose background groups (US1,US2),
gravitationally unbound to the cluster. Significant
small-scales structures are present within the cluster: a triplet (S1),
a close septet (S2) and a very condensed core (C).
 In this case the identification of
the main system  within the main peak  produces macroscopic results. The
cluster becomes regular and the VDP becomes flat in the external
region as expected in a cluster which is in a state of dynamical
equilibrium.
The core has a low velocity dispersion. By subtracting the core structure,
the $\sigma$ of the cluster increases by about 100 \ks, approaching the
observed value of the X-ray temperature.

{\em  A262. ---} True cluster, poorly populated after removing loose
disperse galaxies of the field, whose presence is due to the fact that
this cluster belongs to the Perseus supercluster.

{\em  A399-401. ---} Bimodal system, the two populations (MS1=A401,
MS2=A399) slightly overlapping but separated in velocity by almost 700
\ks. No further substructures.  The separation of A399 and A401 is
considerably difficult because the clusters are fairly close together
in radial velocity (see e.g. Oegerle \& Hill 1994; Girardi et
al. 1996).  The two-body model confirms that these two clusters are
probably gravitationally bound (see also Oegerle \& Hill 1994).  Some
evidence that A401 is a multiple cluster comes from Slezak et
al. (1994).  Recent results by Fujita et al. (1996), based on
X-ray data, suggest that these clusters are really interacting but
that the interaction is not strong at present; however, they cannot
exclude the possibility 
that there was a past first encounter.  As a possible sign of a
substructure, we find that the cD galaxy in A401 cluster has a relevant
peculiar velocity.

{\em A426. ---} Initially extended field. The main cluster (MS) shows
an irregular (elongated) shape and a regular velocity distribution.
There is a strong clustering (C) in the central region. The C
structure appears rather dynamically perturbed (i.e. with a high
dispersion), although its mean velocity well agrees with that of the
cluster.  Indeed, it has been recently claimed that this cluster does
not appear to be in a complete relaxed state.  In particular, Mohr,
Fabricant, \& Geller (1993) found substructures in the core. Also
Slezak et al. (1994) found a double peak in the core by analyzing
X-ray data: however, the region they analyzed is smaller than our
minimum scale analyzed.  This cluster is well-known for showing a
$\beta$-problem ($\beta=1.78^{+0.48}_{-0.34}$ in Edge \& Stewart
1991b).  The $\sigma$ is now in acceptable agreement with the estimate
of $T$ ($\beta=1.25^{+0.24}_{-0.22}$).  Other observational evidence
for reducing the value of $\sigma$ comes also from Fadda et al. (1996,
$\beta=1.01^{+0.24}_{-0.16}$).  The strong increase in the VDP towards
the cluster center suggests the presence of galaxies with radial
orbits in the external cluster region, as already suggested by, e.g.,
Solanes \& Salvador-Sol\'e (1990).  The acceptable agreement between
$\sigma$ and $T$ suggest that these galaxies, although recently
infalled into the cluster, are already roughly virialized within the
cluster potential.

{\em  A539. ---} Two systems (MS1, MS2), both very extended, separated in
velocity by over 4000 Km/s, but spatially overlapped.  MS1, whose
condensed core should be the real structure, has to be considered a
foreground group (see also Pisani 1993).  Cluster A539 should be
the condensed core of the MS2 clump.  In order to better analyze the
cluster we considered the whole published data sample within 6 \h from
the X-ray center of the cluster (A539*).  This sample is not nominally
complete; however, we can probably rely on some uniformity in a small
region, e.g.  the region close to the cluster center, where
substructures are detected.  At the intermediate scale we detected a
core structure C, which contains two structures (C1, C2) at the small scale.
The C1 clump corresponds to the core of the above detected
MS2, while the C2 clump is at higher redshift.

{\em  A548. ---} The very irregular velocity distribution suggests a
complex dynamical status (see e.g. Davis et al. 1995).  The optical
data clearly show a bimodal aspect with two systems, MS1 and MS2, which
correspond respectively to the SW and NE X-ray components (i.e. S2 and
S1 in Davis et al., 1995).  Within the MS1 component we detect 3
substructures (S1,S2,S3) and a core structure (C), which well
corresponds to the X-ray center.  Also Davis et al. (1995), by using
partially different redshift data, found that the SW optical region is
complex.  The $\sigma$ of the MS2 sample well agrees with the respective
estimate of $T$.
On the contrary, the $T$ of the other component differs (but not
significatively within the errors) by about 200 \ks from the $\sigma$
of MS1, but well agrees with the $\sigma$ of its core.  This suggests
that the core is the virialized part of the SW component and is
responsible for the observed X-ray emission. The surrounding galaxies
and clumps may not yet 
be in dynamical equilibrium.  More precise
$T$ estimates could   easily solve this problem.

{\em  A754. ---} The MS sample is elongated and includes two lobes,
while US is an external structure.  At the intermediate scale, we
detected the MS1 (NW) and the MS2 (SE) lobes; C is a core structure
detected in MS2 and represents half of the lobe. D88 did not detect any significant substructure, but bimodality
is displayed in recent X-ray and optical data (e.g.  Zabludoff \&
Zaritsky 1995).  In the VDP (see Figure~4) we used the galaxy density
centers rather than the X-ray one.  In fact, optical structures do not
correspond to the X-ray ones in this cluster, which shows direct
evidence of an on-going collision (see Henry \& Briel 1995; Zabludoff
\& Zaritsky 1995; Heriksen \& Markevitch 1996).

{\em A1060. ---} Almost regular cluster with a dynamically perturbed
condensed core (C) and a close quintet at South (S1).  The core
structure appears better defined by using the less complete sample
(A1060*) whose completeness level is only 50\%.  In this second sample
the core structure represents an important fraction of the total
population of the cluster.  Although this cluster appears very
regular, several authors have suggested the presence of substructures.
For instance, Fitchett \& Webster (1987) pointed out that their
$\sigma$ value is too high to agree with the value expected from X-ray
luminosity.  Indeed, our $\sigma$, which is similar to their value, is
too low with respect to the X-ray temperature.  The apparent
discrepancy is explained by the fact that this cluster does not fit
the usual relation between luminosity and temperature (e.g. David et
al. 1993).  This finding could suggest some anomalies in the dynamical
status of the gas, rather than in that of the galaxies.

{\em  A1146. ---} Distant and regular cluster.  The agreement with 
$T$ is sufficiently good if we consider that this $T$ is not measured
but only estimated from  X-ray luminosity.

{\em  A1185. ---} The field usually attributed to A1185 does not refer
to the cluster itself, since it includes a large and uniform
environment.  The true cluster could consist in the condensed
structure in the center (C).  Already Fadda et al. (1996) noted the
anomalous increase of the VDP in the external region of the cluster
and naively suggested neglecting this region.

{\em  A1367. ---} This cluster belongs to the Coma supercluster.
Irregular in shape and velocity distribution, this cluster contains a
close sextet (S).

{\em A1631. ---} Very irregular in shape despite a regular velocity
distribution. This cluster shows a very elongated structure (C), which
passes through the cluster center and contains some bright
galaxies. This anomalous shape could suggest that the real core
structure might be C1, which is a close significant sextet included in
C.  These central structures were not detected by D88. Moreover, the
substructures found by E94, who analyzed the same galaxy sample, are
not detected in this work; this discrepancy with E94 is discussed in
\S~3.1.

{\em  A1644. ---} Regular cluster. The velocity distribution 
shows a secondary peak which corresponds to an extended foreground
structure ($\Delta$V = 1100 Km/s).  Both populations are fully
overlapping with each other and cannot be separated; no significant US
structure is identified.  We use the same galaxy sample as Dressler \&
Schechtman (1988b), who found that the presence of substructures in
this cluster is not definitively significant (c.l. about 97\%).  E94
found substructures which are not shown here.  However, the peculiar
velocity of the cD galaxy suggests the possible presence of a minor
substructure.

{\em A1736. ---} Cluster A1736 shows two well separated peaks in the
velocity distribution. The main peak corresponds to A1736B, according
to the denomination by D88. We do not analyze the foreground peak
(A1736A) because of the small number of objects.  By analyzing the
same data sample, D88 found substructures, which are significant only
at about 98\%. Fadda et al. (1996) found two peaks in the velocity
distribution, but so strongly overlapping with each other that their
physical separation was uncertain.  According to our analysis, A1736B
does not show substructures and this finding is confirmed by the good
agreement with $T$.

{\em  A1795. ---} Regular cluster with a central condensation (C).
Hill \& Oegerle (1993) detected substructures significant only at
about 97\%.  The VDP of the main sample strongly decreases towards the
cluster center.  The low-$\sigma$ core structure C contains a cD
galaxy, whose velocity is consistent with the mean velocity of the C
structure and that of the whole cluster.  Moreover, a
strong cooling flow was observed 
(Edge et al. 1992; Cardiel, Gorgas \& Aragon-Salamanca 1995).  All
these findings indicate that this cluster is a very relaxed cluster
(see also the discussion in \S~5.3).
  The better agreement of $T$ with the $\sigma$ of the MS-C sample rather
than with the MS sample could confirm that the galaxies in the core could
have been slowed down by relaxation processes (e.g. dynamical
friction). 

{\em A1809. ---} Poor cluster, no substructures.  Oegerle \& Hill
(1994) found no presence of substructures either.  The VDP referred to
the X-ray center is very noisy in the center; thus we prefer to show
the VDP referred to the galaxy density center, which shows a clear
decrease towards the central region.

{\em A1983. ---} This cluster is characterized by an asymmetric
velocity distribution, but this irregularity vanishes after removing a
significant foreground quintet located close to the center (S).  The
situation is similar to that found in A85, where the S could be an
unbound foreground group, but in this case the S substructure causes a
less significant variation of the total $\sigma$.  D88 found the
presence of substructures only significant at about 94\%. E94 found a
substructure both in the center and in the cluster field, but none
associated with the S detected in this work.

{\em  A2052. ---} Irregular cluster, contains some disturbing objects
(pairs, triplets) which, however,  do not constitute  any significant
substructure.  Malumuth et al. (1992) found no significant structure,
either.  The global $\sigma$ is lower than $T$, but agrees within two
s.d..

{\em A2063. ---} This field contains two clusters that are very
distant from each other in terms of velocity ($\Delta$V $\sim$ 3000
Km/s): MS1 and MS2, which correspond to A2063 and MKW3S, respectively.
The MS1 sample contains a substructure of seven very close galaxies
(S), largely background. However, according to the Newton criterion,
we cannot exclude the possibility that S is bound to the remaining
galaxies in the MS1 sample. For the same reason we cannot rule out the
possibility that MS1 and MS2 samples may be bound.  Both clusters
contain cD galaxies; however, the cD galaxy of A2063 has a peculiar
velocity: this peculiarity disappears when we reject its substructure
S.

{\em A2107. ---} Remarkably regular cluster.  Oegerle \& Hill (1992)
found evidence of substructures by using the test of D88.  The
peculiar velocity of the cD galaxy could suggest a situation of
non-perfect dynamical equilibrium.  However, with the present data,
there is good agreement between $\sigma$ and $T$.

{\em  A2124. ---} Regular cluster.

{\em  A2151. ---} Highly structured cluster, despite  an apparently
regular velocity distribution of the main field.  We notice a
significant central structure C ($\Delta V \sim -700$ \ks), an extended
subsystem S1 at North and a close background quintet S2 at East. These
three subsystems, very distant from each other, concern about 50\% of the
total population of the cluster.  These results are consistent with
E94. Bird, Davis, \& Beers (1995) find that X-ray and optical
distributions are not very similar. We confirm this result: in
particular, our central C is not centered on the X-ray cluster center.
This fact and the large velocity of C with respect to that of the
whole cluster suggest that C could be considered a real substructure
rather than a particularly relaxed central region. This cluster should
be regarded as a case of present cluster merging.  Moreover, although
the $\sigma$ of the MS sample is in good agreement with the $T$ by David
et al. (1993), it is higher than the $T$s detected by Bird et
al. (1995) for the individual clumps.

{\em  A2197-2199. ---} No significant structures, thought elongated in
shape.  The field we analyzed contains, however, the two well-known
clusters A2197 and A2199, which  correspond closely to the structures we
detect at a very low significance level (about 60\%).  The background
system MS1 in the South (A2199) contains a close foreground quartet
(S1), while the foreground system MS2 at North (A2197) contains a
background quartet (S2). This peculiar situation is very difficult to
analyze with our method, since the local kinematics cannot be cleaned
from the mutual contamination produced by the four systems.
Therefore, we do not consider these clusters in the final discussion.

{\em A2634-2666. ---} The MS sample includes both A2634 and the tiny
cluster A2666 (S1) and a further substructure (S2). We identify A2634
with the MS-S1-S2 sample, i.e. the MS sample after the rejection of
A2666 and S2, which are both unbound to the remaining galaxies.  The
``identified cluster'' main properties are: Center 1950
($\alpha$,$\delta$)=233606.9+264541; $N$=216; $R_{max}$=5.64;
$\overline{v}=9136$ \ks; $P_W<0.1$\%; $\sigma$=886.  Scodeggio et
al. (1995) did not find any evidence of substructures in the central
cluster region.  The $\sigma$ of the MS-S1-S2 sample is in good
agreement with the $T$ at about 1 \h, and an increase in the outer
region is likely because of the presence of some remaining
interlopers.  The $T$ of A2666 is much higher than our value of
$\sigma$, but we note that this $T$ is estimated from X-ray
luminosity.

{\em A2670. ---} We analyze two samples. A2670, the first sample, does
not contain any substructures. The second sample A2670$^*$, which is
deeper and has a smaller extension, contains a foreground group of
galaxies (US) at West of the main system. Sharples et al.
(1988) did not find any firm statistical evidence of subclustering,
either. On the contrary, E94 found a series of structures which are
not present in this analysis.  B95, by taking into account the
presence of substructures, reduces the peculiar velocity of cD. Here
we still found a peculiar velocity for the cD galaxy, but this
peculiarity disappears when we consider the deepest sample A2670$^*$.

{\em A2717. ---} This cluster contains a significant core structure
(CS), slightly foregrounded ($\Delta$V $\sim$ -600 \ks.), which
involves about 40\% of the whole population and contains the cD
galaxy. This feature is responsible for the asymmetry observed in the
velocity distribution.  The cD galaxy, which has not a peculiar
velocity with respect to the MS sample but only with respect to the C
structure, confirms that this core is probably dynamically perturbed.

{\em  A2721. ---} Regular cluster.  We show the VDP referred
to the galaxy density center, which is more regular than that computed 
with the X-ray center.

{\em  A2877. ---} Regular velocity distribution, irregular shape.  We
detect two significant subsystems: the core structure (C), and a
structure (S) at North.

{\em  A3128. ---} The analyzed field contains an extended structure at
South (US), which gives the cluster its elongated shape. The main
cluster MS is regular, with a very condensed structure in the center
(C).  We report the VDP computed with the galaxy density center, which
is more regular than that referred to the X-ray center.

{\em  A3266. ---} The analyzed field contains two systems, which are
very similar in terms of velocity but spatially far apart. The
smallest group is elongated (US) and lies at East of the main dominant
system (MS). This one is regular and should correspond to the
real cluster.  The VDP of MS is very noisy in the central region and
the $\sigma$ is higher than the $T$, but consistent within two s.d.. 

{\em A3376. ---} Regular velocity distribution despite its irregular
shape. The central region appears elongated.  The South part of this
group appears well coincident with the X-ray center; thus we prefer to
denominate it a core structure, although it is probable that its North
component may be a real substructure.  D88 found substructures less
significant than 95\%. The substructures detected by E94 are not
present in this analysis (see \S~3.1).

{\em  A3391-3395. ---} Rich field with two distinct clusters, partially
overlapping, but well separated in velocity. The dominant system
(MS1=A3395) is a rich extended cluster, which has a regular velocity
distribution but is almost irregular in shape. The small scale
analysis reveals a significant central condensation (C1), which is
elongated and dynamically similar to the main system. A small system
at North (MS2=A3391), in the background of MS1 ($\Delta$V $\sim$ 2000
\ks), is formed by a core structure (C2) appended to some loose
background galaxies.  Girardi et al. (1996) pointed out the difficulty
in separating the two clusters and used the VDP to truncate the
clusters at the radius, where the VDP increases owing to the presence of a
close cluster.  The present method is able to separate the clusters
and gives VDPs, which are flat in the external region. The cD galaxy
of A3391 has a peculiar velocity with respect to MS2, but not with
respect to C2, which contains the cD galaxy.

{\em A3526. ---} We analyzed two samples (A3526 and A3526*), the
former deeper in magnitude and with a minor extension than the latter.
In both samples we identify two dominant significant systems (MS1,
MS2). They are fairly well separated in velocity ($\Delta V \sim 1500$
\ks), but fully overlapped and probably bound.  MS2 contains a
condensed structure at East (C), whose center corresponds to its
galaxy density center.  In the sample A3526$^*$, MS1 includes a group
of foreground galaxies (S), which is probably unbound ($\Delta$V =
-1000 \ks).  The velocity distribution of MS1 becomes regular if we
exclude S.

Already Lucey, Currie, \& Dickens (1986) detected the presence of two
peaks in the velocity distribution and found that only a minority of
the galaxies (30$\%$) in the secondary peak in the cluster A3526 could
actually be distant from the primary peak. The cluster should be thus
regarded as a strongly substructured cluster (see also Mohr et
al. 1993).  Girardi et al. (1996) and Fadda et al. (1996) indicated
this cluster as one with a problematic dynamics by using the velocity
distribution.  A likely on-going merging might explain the high
temperature of the collision-heated gas.

{\em Shapley region. ---} The analyzed field contains two well sampled
main clusters: A3558 and A3562.  MS is the dominant group, which
contains the cluster A3558.  US1 is a small group at West, clearly
separated in position on the basis of the available data. US2 is a
large group at East, slightly foregrounded with respect to US1.  Both
US1 and US2 have a high dispersion because they probably do not
correspond to true physical systems.  In fact, US2 contains a
structure (US2S), which is significant at the 80\% c.l. and correspond
to A3562. Due to the limited value of significance, we do not consider
this cluster in the final discussion, although the good agreement with
$T$ suggests a good real identification.  MS is a rich and very
structured system despite its regular velocity distribution.  The
analysis at small scales reveals three distinct groups inside: a
background structure S1 at North (+800 \ks); a very condensed
structure at East S2, which also contains the poor cluster SC1329-314;
a central structure S3, which is dynamically similar to the main
system and corresponds to the identified A3558 cluster.

The internal structure of A3558 is widely debated in the literature
and the $T$ of this cluster is well known to be lower than $\sigma$
(see e.g.  by Bardelli et al. 1996, who report $\beta=1.79$).  Our S3
structure, which we identify as A3558, has a $\sigma$ compatible with
$T$ ($\beta=0.86^{+0.39}_{-0.23}$), although the peculiar velocity of
the cD galaxy may suggest the presence of internal substructures.

{\em  A3667. ---} Asymmetric and dynamically regular cluster, although
it includes a significant, very condensed structure S, which is
remarkably elongated. No further substructures appear
significant.  The global $\sigma$s of MS or MS-C samples are in good
agreement with $T$, although the VDP is very noisy in the central
cluster region. 

{\em  A3716. ---} Cluster A3716 is clearly divided into  North and 
South components (D88). Hence, this cluster is an
apparent case of large-scale structure. The MS we detect is the
richest South component, since the data of North component are too
scarce. This South component is an apparently regular structure.
We adopt the optical center for the computation of the VDP, since it is
more regular towards the center than the VDP referred to the X-ray
center.

{\em A3888. ---} Almost regular cluster but with an asymmetric
velocity distribution.  The local dispersions are atypically high due
to the presence of numerous field galaxies uniformly distributed
through the whole extension of the cluster, thus suggesting a
systematic contamination effect.

{\em A4038. ---} Regular cluster.  Unfortunately, the region of this
cluster we studied is too small to permit a good check of the VDP. The
VDP centered on the optical center is less noisy than that centered on
the X-ray center, but we cannot be sure that the VDP remains flat
towards the external cluster regions.


\end{document}